\documentclass[usenatbib,useAMS]{mnras}

\usepackage{graphicx}	
\usepackage{amsmath}	
\usepackage{amssymb}	
\usepackage{rotating}
\newcommand\target{{GRS 1915+105} }

\usepackage[T1]{fontenc}
\usepackage{ae,aecompl}

\usepackage{newtxtext,newtxmath}

\title[Type-C QPO phase lags in GRS 1915+105]{A Systematic Analysis of the Phase Lags Associated with the Type-C Quasi-periodic Oscillation in \target}

\author[Zhang et al.]{Liang Zhang$^{1,2}$\thanks{E-mail: liang.zhang@soton.ac.uk}, 
Mariano M\'{e}ndez$^{3}$,
Diego Altamirano$^{1}$, 
Jinlu Qu$^{2}$, 
Li Chen$^{4}$, 
\newauthor 
Konstantinos Karpouzas$^{1,3}$, 
Tomaso M. Belloni$^{5}$,
Qingcui Bu$^{2,6}$,
Yue Huang$^{2}$,
\newauthor
Xiang Ma$^{2}$,
Lian Tao$^{2}$
and
Yanan Wang$^{7}$
\\
$^{1}$Physics and Astronomy, University of Southampton, Southampton, Hampshire SO17 1BJ, UK\\
$^{2}$Key Laboratory for Particle Astrophysics, Institute of High Energy Physics, CAS, Beijing 100049, China\\
$^{3}$Kapteyn Astronomical Institute, University of Groningen, PO Box 800, NL-9700 AV Groningen, the Netherlands\\
$^{4}$Department of Astronomy, Beijing Normal University, Beijing 100875, China\\
$^{5}$INAF-Osservatorio Astronomico di Brera, via E. Bianchi 46, I-23807 Merate, Italy\\
$^{6}$Institut f\"ur Astronomie und Astrophysik, Kepler Center for Astro and Particle Physics, Eberhard Karls Universit\"at, Sand 1, 72076 T\"ubingen, Germany\\
$^{7}$Universit\'e de Strasbourg, CNRS, Observatoire astronomique de Strasbourg, UMR 7550, F-67000 Strasbourg, France
}
\date{Accepted 2020 March 18. Received 2020 March 16; in original form 2019 December 18}

\pubyear{2020}

\begin{document}
\label{firstpage}
\pagerange{\pageref{firstpage}--\pageref{lastpage}}
\maketitle

\begin{abstract}
We present a systematic analysis of the phase lags associated with the type-C QPOs in GRS 1915+105 using {\it RXTE} data. Our sample comprises of 620 {\it RXTE} observations with type-C QPOs ranging from $\sim0.4$ Hz to $\sim6.3$ Hz. Based on our analysis, we confirm that the QPO phase lags decrease with QPO frequency, and change sign from positive to negative at a QPO frequency of $\sim2$ Hz. In addition, we find that the slope of this relation is significantly different between QPOs below and above 2 Hz. The relation between the QPO lags and QPO rms can be well fitted with a broken line: as the QPO lags go from negative to positive, the QPO rms first increases, reaching its maximum at around zero lag, and then decreases. The phase-lag behaviour of the subharmonic of the QPO is similar to that of the QPO fundamental, where the subharmonic lags decrease with subharmonic frequency and change sign from positive to negative at a subharmonic frequency of $\sim1$ Hz; on the contrary, the second harmonic of the QPO shows a quite different phase-lag behaviour, where all the second harmonics show hard lags that remain more or less constant. For both the QPO and its (sub)harmonics, the slope of the lag-energy spectra shows a similar evolution with frequency as the average phase lags. This suggests that the lag-energy spectra drives the average phase lags. We discuss the possibility for the change in lag sign, and the physical origin of the QPO lags. 
\end{abstract}


\section{Introduction}\label{sec:intro}

Fast X-ray variability is an important characteristic of black hole binaries (BHBs), and a key to understand the physical processes related to accretion in these systems. Power density spectra (PDS) is a commonly used tool to study these variability \citep{Klis1989}. In a PDS of BHBs, the most prominent feature is one or more narrow peaks, which are known as quasi-periodic oscillation (QPO; see \citealt{Klis2006}). As these QPOs are thought to arise from the innermost regions of the accretion flow, they can be used to probe the effect of General Relativity around black holes \citep[e.g.][]{Belloni2019}. Based on the frequency range, QPOs in BHBs are normally divided into two groups: low-frequency QPOs with frequencies ranging from a few mHz to $\sim 30$ Hz, and high-frequency QPOs with frequencies up to $\sim 500$ Hz (see \citealt{Motta2016} for a recent review). Low-frequency QPOs have been observed in almost all the transient BHBs \citep[e.g.][]{Miyamoto1991,Takizawa1997,Mendez1998,Tomsick2000,Kalemci2003,Casella2004,Homan2005,Yadav2016,Huang2018,Xu2019}, and can be further classified into three categories, dubbed type-A, -B, and -C, based on differences in PDS shape and phase-lag behaviour \citep{Wijnands1999,Remillard2002,Casella2005}. In contrast, only nine sources show a handful of high-frequency QPOs \citep[e.g.][]{Morgan1997,Miller2001,Strohmayer2001,Homan2003,Altamirano2012,Belloni2012,Mendez2013}. In this paper, we will focus our analysis on the so-called type-C QPOs. A similar analysis of type-B QPOs will be presented in a separate work. Type-A QPOs are not considered due to the small number of detections.

Type-C QPOs are the most common type of QPOs in BHBs; these QPOs show in the PDS as a strong, narrow peak with variable frequency, superposed on a flat-top broad-band noise component. A subharmonic and a second harmonic peak are often present in the PDS \citep[e.g.][]{Casella2005}. Although a number of models have been proposed to explain type-C QPOs, their physical origin is still under debate. These models either consider instabilities in the accretion flow \citep[e.g.][]{Tagger1999,Cabanac2010}, or geometrical effects, such as Lense-Thirring precession \citep{Stella1998,Schnittman2006,Ingram2009}. Recently, \citet{Heil2015} and \citet{Motta2015} found evidence that the amplitude of the type-C QPOs depends on the inclination of the accretion disc with respect to the line of sight, consistent with the prediction in the precessing ring model proposed by \citet{Schnittman2006}. In addition, the iron line equivalent width \citep{Ingram2015}, centroid energy \citep{Ingram2016}, and the reflection fraction \citep{Ingram2017} have been found to be modulated with QPO phase. This can be interpreted as the inner flow illuminating different azimuths of the accretion disc as it precesses, causing the iron line to rock between red- and blue-shift. All of these results suggest a geometric origin of the type-C QPOs, with Lense-Thirring precession being the most promising model \citep{Ingram2009}. 

Phase/time lag between different energy bands is another powerful tool to study fast X-ray variability. Recently, \citet{Eijnden2017} found that the phase lags at the type-C QPO frequency also depend on inclination: at low frequencies, all sources except \target tend to show zero phase lags; at high frequencies, low-inclination (face-on) sources display hard lags, while high-inclination sources (edge-on) display soft lags. Such an inclination dependent phase-lag behaviour also points to a geometric QPO origin. The relation between the QPO phase lags and the QPO frequency is found to differ between sources \citep{Eijnden2017}. \citet{Zhang2017} studied this relation in GX 339$-$4 in detail. The clear break found in the phase-lags vs. frequency relation at around 2 Hz suggests that different mechanisms may be responsible for the QPO lags at different phases of the outburst. Another interesting result related to phase-lag behaviour is that the QPO fundamental and its (sub)harmonics exhibit lags of different signs. In general, the second harmonic always shows a hard lag, while the subharmonic usually shows a soft lag \citep{Casella2005,Eijnden2017}. 

\target has been active since its discovery in 1992 \citep{Castro1992} and has become one of the best studied Galactic BHBs. The source shows complex timing and spectral properties that are quite different from other BHBs (except for IGR J17091$-$3624, \citealt{Altamirano2011}). The X-ray variability can be classified into approximately 14 separate classes, based on the properties of its light curves and colour-colour diagrams \citep{Belloni2000,Klein2002}. Each of these variability patterns can be further reduced to transitions between three basic states \citep{Belloni2000}: a hard state, and two soft states with different luminosities. Many instances of type-C QPOs with frequencies ranging from 0.1 Hz to 10 Hz have been observed in the hard state \citep[e.g.][]{Morgan1997,Reig2000,Qu2010,Pahari2013,Yadav2016}. The observed hard rms spectra of these QPOs indicate a possible origin of the QPOs in the corona \citep{Yan2013}. The QPO frequency is found to be energy dependent \citep{Qu2010,Yan2012}, which can be interpreted as the result of differential precession of the hot inner flow \citep{Eijnden2017}.

The phase lags of the type-C QPOs in \target have been studied by several previous work, such as \citet{Reig2000}, \citet{Qu2010}, \citet{Pahari2013}, and \citet{Eijnden2016}. The system exhibits a very complex phase-lag behaviour: as QPO frequency increases, the phase lags at the frequencies of the QPO decrease and change sign from positive to negative at around 2 Hz \citep{Reig2000,Qu2010,Pahari2013}. The QPO lags also show energy dependence, and sometimes a break at $6-7$ keV is present in the lag-energy spectra \citep{Pahari2013}. For the second harmonic, \citet{Reig2000} reported that the lags change sign from positive to negative at around 4 Hz. But \citet{Pahari2013} found that the second-harmonic lags are always positive, and remain roughly constant. There are no previous studies on the evolution of the subharmonic lags so far. 

It is worth noting that all the results above related to phase lags in \target are based on a small sample of observations. For example, \citet{Reig2000} and \citet{Qu2010} used 43 and 19 \emph{RXTE} observation intervals, respectively, to study the frequency- and energy-dependent type-C QPO phase lags. \citet{Pahari2013} used 12 \emph{RXTE} observations to study the difference between the lags in the radio-loud (plateau) state and the radio-quiet state. Thus, some of the previous results might not be representative, such as the evolution of the (sub)harmonic lags, and the properties of their energy dependence. Furthermore, the mechanism producing these lags is still not fully understood. Therefore, in this paper, we present a systematic analysis of the phase lags associated with the type-C QPOs in \target using {\it RXTE} data. We measured the phase lags at the frequencies of the QPO and its (sub)harmonics, and studied in detail their dependence on frequency and photon energy. We describe our sample and data analysis methods in Section \ref{sec:obs}. We present our results in Section \ref{sec:results} and discuss them in Section \ref{sec:Discussion}.

\section{Observations and data analysis} \label{sec:obs}

We examined all the \emph{RXTE} archival observations (1800+) of \target spanning from 1996 to 2012. For the timing analysis, we used the software GHATS version 1.1.1 under IDL\footnote{GHATS, \url{http://www.brera.inaf.it/utenti/belloni/GHATS_Package/Home.html}}. Binned, event and GoodXenon data modes were used. For each observation, we computed an averaged PDS in the full energy band (absolute PCA channel $0-249$) using 128 s long intervals and 1/128 s time resolution, corresponding to a Nyquist frequency of 64 Hz. The PDS were normalized according to \citet{Leahy1983}, and the contribution due to Poisson noise was subtracted \citep{Zhang1995} in GHATS. A logarithmic rebin was applied to the PDS, while the size of a bin increases by exp(1/100) with respect to the size of the previous one.

\begin{figure}
\resizebox{\columnwidth}{!}{\rotatebox{0}{\includegraphics{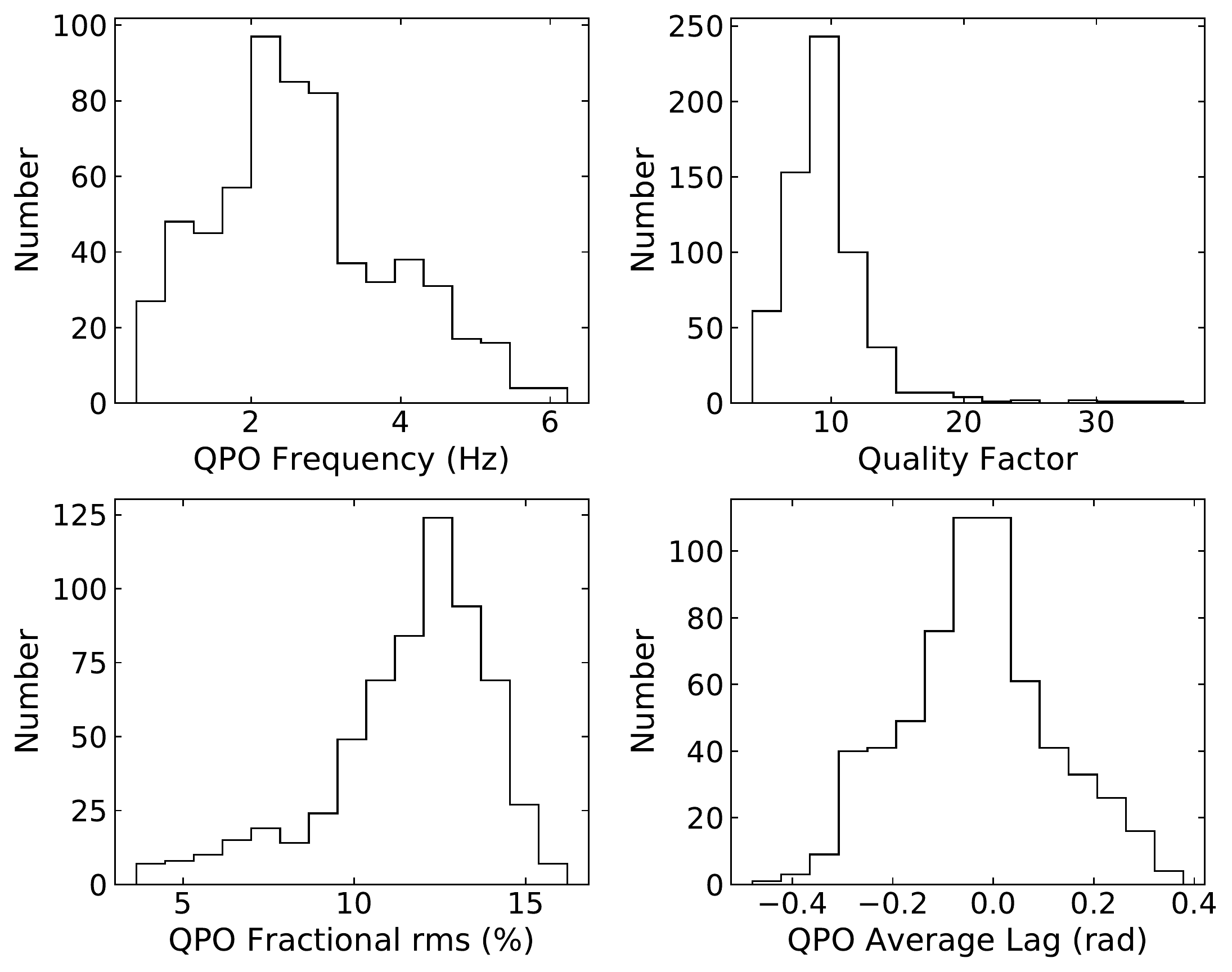}}}
\caption{The distribution of the centroid frequency, quality factor, fractional rms amplitude in the full PCA band, and average phase lag between photons in the 2--5.7 keV and 5.7-15 keV bands for the type-C QPOs in GRS 1915+105.}
\label{fig:hist}
\end{figure}

\begin{figure*}
\centering
\resizebox{2\columnwidth}{!}{\rotatebox{0}{\includegraphics{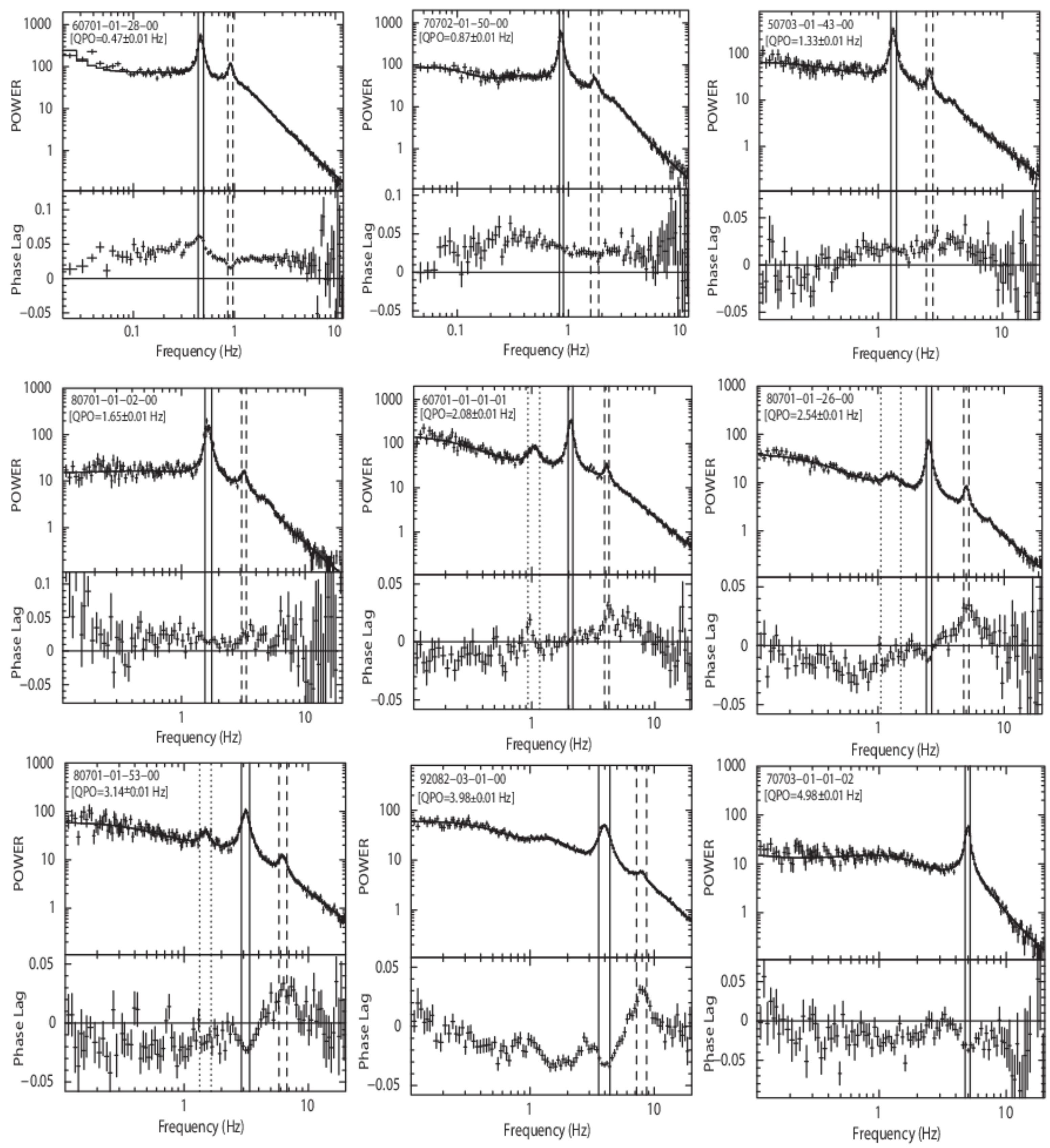}}}
\caption{Examples of power and lag-frequency spectra for nine observations of \target with the type-C QPOs at different centroid frequencies. The power spectra were calculated for the full PCA band, while the phase lags were computed between the $2-5.7$ keV and $5.7-15$ keV bands. A positive lag means that hard photons lag the soft ones. All these observations are from PCA Epoch 5. Observation ID and QPO fundamental frequency are shown in the top left corner of each panel. The vertical, dashed and dotted lines indicate the ranges over which the QPO fundamental, second-harmonic and subharmonic lags are averaged ($\nu_{0} \pm \rm{FWHM}/2$), respectively.}
\label{fig:pds}
\end{figure*}

For our analysis we selected only observations where at least one narrow QPO peak is present on top of the broadband noise component in the PDS, which is typical for the type-C QPOs. Following \citet{Belloni2002}, we fitted these PDS with a sum of Lorentzian functions using XSPEC version 12.9. Based on the fitting results, we excluded features with a significance\footnote{The significance of QPOs is given as the ratio of the integral of the power of the Lorentzian used to fit the QPO divided by the negative 1$\sigma$ error on the integral of the power.} of less than $3\sigma$ or a $Q$ factor\footnote{$Q=\nu_{0}/{\rm FWHM}$, where $\nu_{0}$ is the centroid frequency of the Lorentzian component and FWHM is its full width at half maximum.} of less than 2. We further excluded $\sim90$ observations in which the QPO frequency changes significantly with source intensity by checking the dynamical power spectra. Our final sample includes a total of 620 observations. 

\begin{figure*}
\resizebox{2\columnwidth}{!}{\rotatebox{0}{\includegraphics{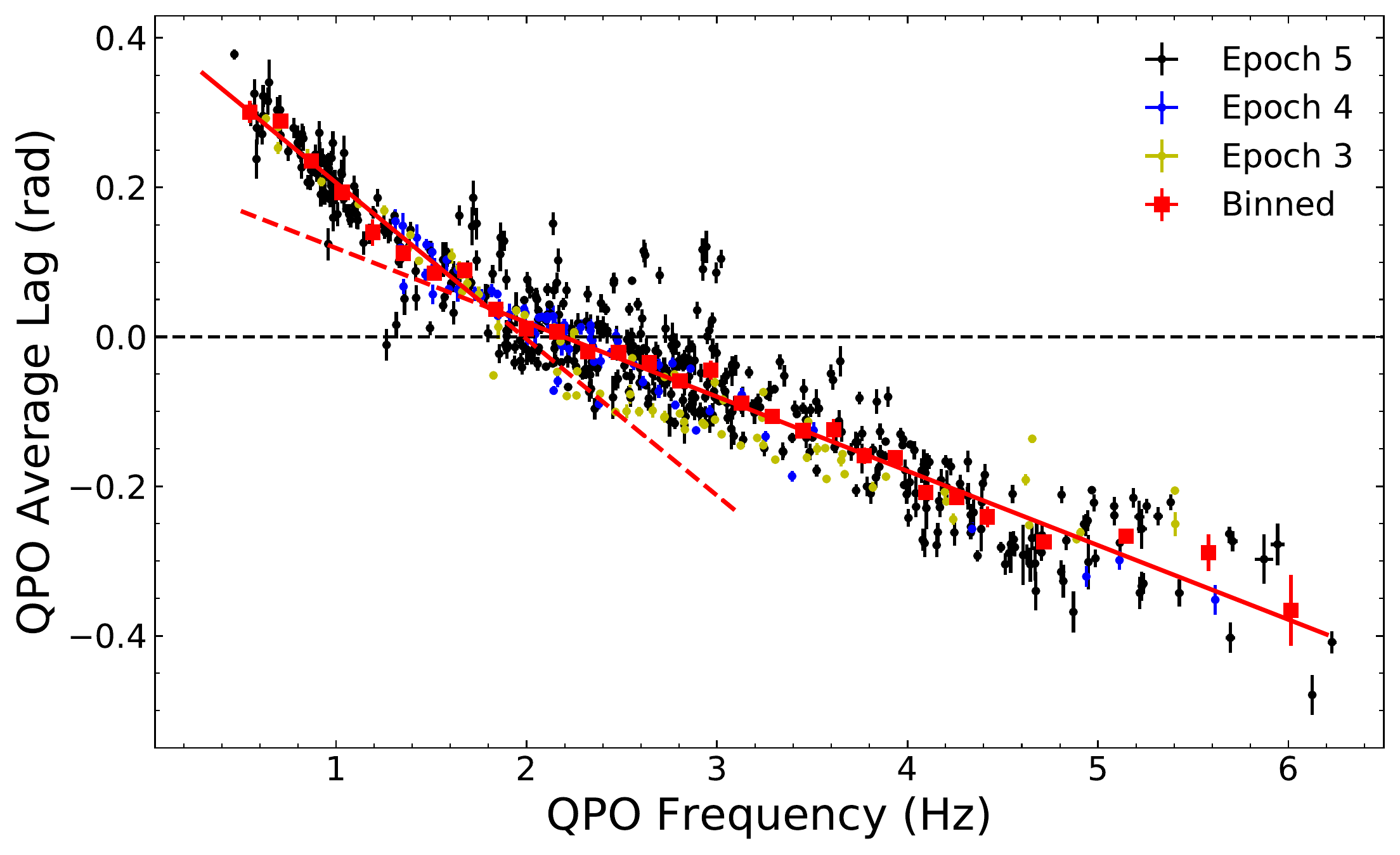}}}
\caption{Average phase lags as a function of the QPO frequency for the fundamental of the type-C QPO in GRS 1915+105. The solid lines indicate the best-fitting broken line to the binned data. The dashed lines are the continuation of the broken line, drawn for comparison with the data in those parts of the plot.}
\label{fig:fre_phase}
\end{figure*}

Following the method described in \citet{Vaughan1997} and \citet{Nowak1999}, we produced a frequency-dependent phase lag spectrum (lag-frequency spectra) between the $2-5.7$ keV and $5.7-15$ keV energy bands for each observation. Our final sample includes observations during the PCA instrument Epochs $3-5$. To account for changes in the PCA gain\footnote{See \url{https://heasarc.gsfc.nasa.gov/docs/xte/e-c_table.html}}, we selected the absolute channels most approximately matching these energy bands, but the exact energy bands still differ slightly between Epochs. To calculate the phase lags at the frequencies of the QPO and its (sub)harmonics, we averaged the phase lags over the width of each Fourier component, around its centroid frequency, $\nu_{0} \pm \rm{FWHM}/2$. Note that the calculated phase lags at the frequencies of the QPO and its (sub)harmonics are affected by the lags of the broad-band noise component. However, \citet{Eijnden2016} have shown that in GRS 1915+105, the phase lags at the frequency of the QPO are dominated by the QPO itself and the contribution of the noise component is negligible. In this work, a positive lag means that the hard photons lag the soft photons. No correction for the dead-time driven cross-talk effect \citep{Klis1987} was done because this effect was found to be negligible. 

We also calculated the energy-dependent phase lags for both the QPO and its (sub)harmonics (lag-energy spectra), following the procedure described in \citet{Uttley2014}. The phase lags were calculated for the energy bands approximately $4-6$ keV, $6-8$ keV, $8-11$ keV, $11-15$ keV, $15-21$ keV, and $21-44$ keV, with reference to the softest band, $2-4$ keV. In 18 of the 620 observations, it was not possible to carry out the energy-dependent analysis due to the format of the binned data. Because of the small amplitudes of the (sub)harmonics compared to the fundamental, at high energies, their phase lags show large uncertainties. Therefore, in these cases, we ignore the data above 20 keV. 

Finally, we calculated the fractional rms in the full PCA band for the QPO and its (sub)harmonics. We also calculated the energy-dependent fractional rms of the QPO (QPO rms spectra). For this purpose, we calculated the QPO fractional rms in the $2-6$ keV, $6-8$ keV, $8-11$ keV, $11-15$ keV, $15-21$ keV, and $21-44$ keV energy bands. 

\begin{table}
\caption{Properties of the type-C QPO fundamental and its (sub)harmonics in GRS 1915+105.\label{tab:QPO property}}
\begin{tabular}{lccc}
\hline\hline
Property & Fundamental & Second Harmonic & Subharmonic\\
\hline
Frequency (Hz) & $\sim 0.4-6.3$ & $\sim 0.9-9.0$ & $\sim 0.6-2.1$\\
Ratio$^a$      &        -       & $1.86-2.07$    &   $0.43-0.54$ \\
$Q$ ($\nu_{0}$/FWHM)  & $4-37$ & $3-30$ & $2-10$ \\
rms$^b$ (\%) & $3.6-16.2$ & $1.6-8.2$ & $2.6-7.6$ \\
\hline
\multicolumn{4}{l}{$^a$ The frequency ratio between the (sub)harmonic and the fundamental.}\\
\multicolumn{4}{l}{$^b$ The fractional rms amplitude was calculated over the full \emph{RXTE} band.}
\end{tabular}
\end{table}

\section{Results} \label{sec:results}

\begin{figure*}
\centering
\resizebox{2\columnwidth}{!}{\rotatebox{0}{\includegraphics{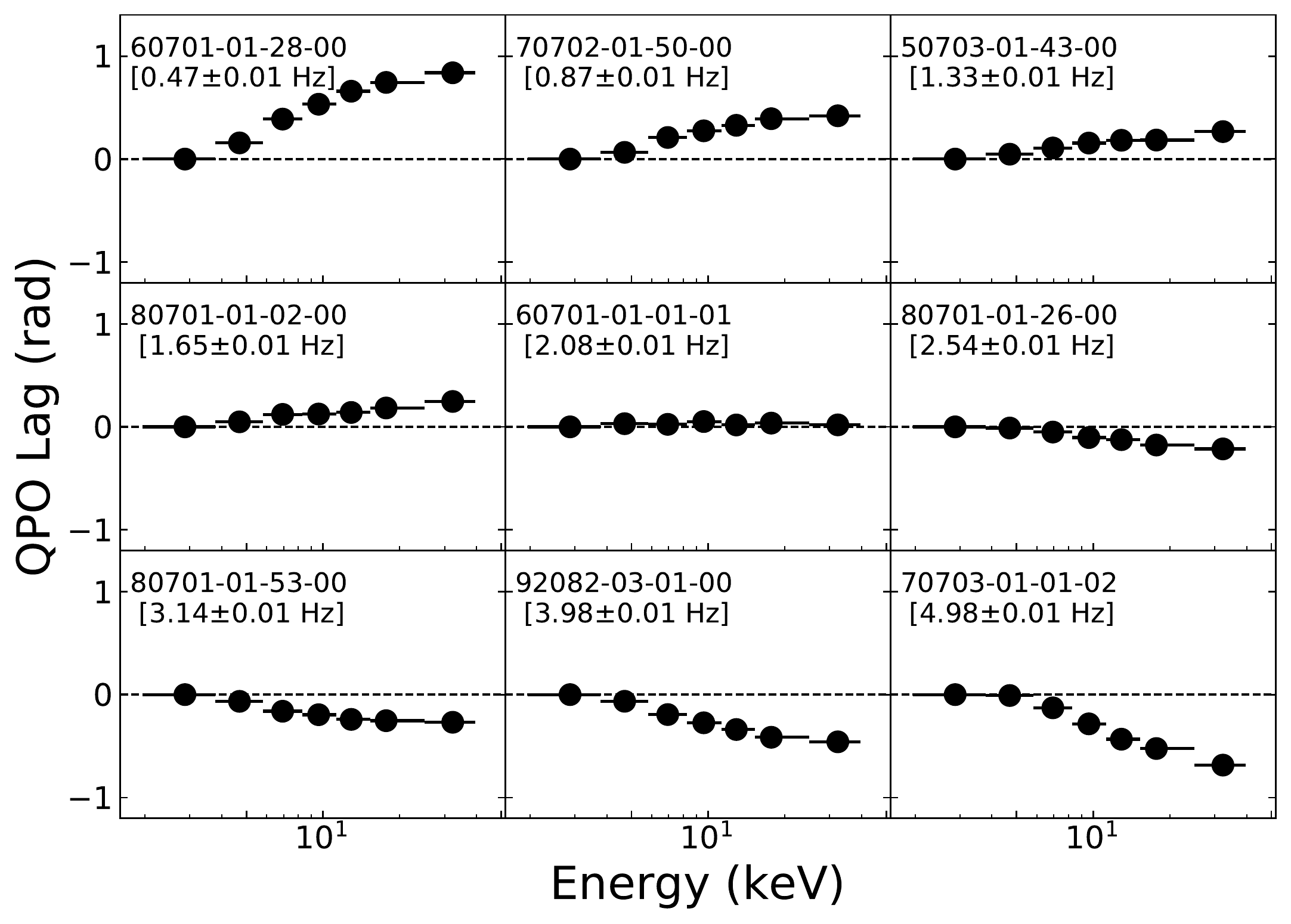}}}
\caption{Representative examples of the lag-energy spectra for the fundamental of the type-C QPO in GRS 1915+105. The observations we show here are the same ones as in Fig. \ref{fig:pds}. Observation ID and QPO frequency are listed in each panel. The reference energy band is indicated with zero phase lag.}
\label{fig:phase_spectra}
\end{figure*}

\begin{figure}
\centering
\resizebox{\columnwidth}{!}{\rotatebox{0}{\includegraphics{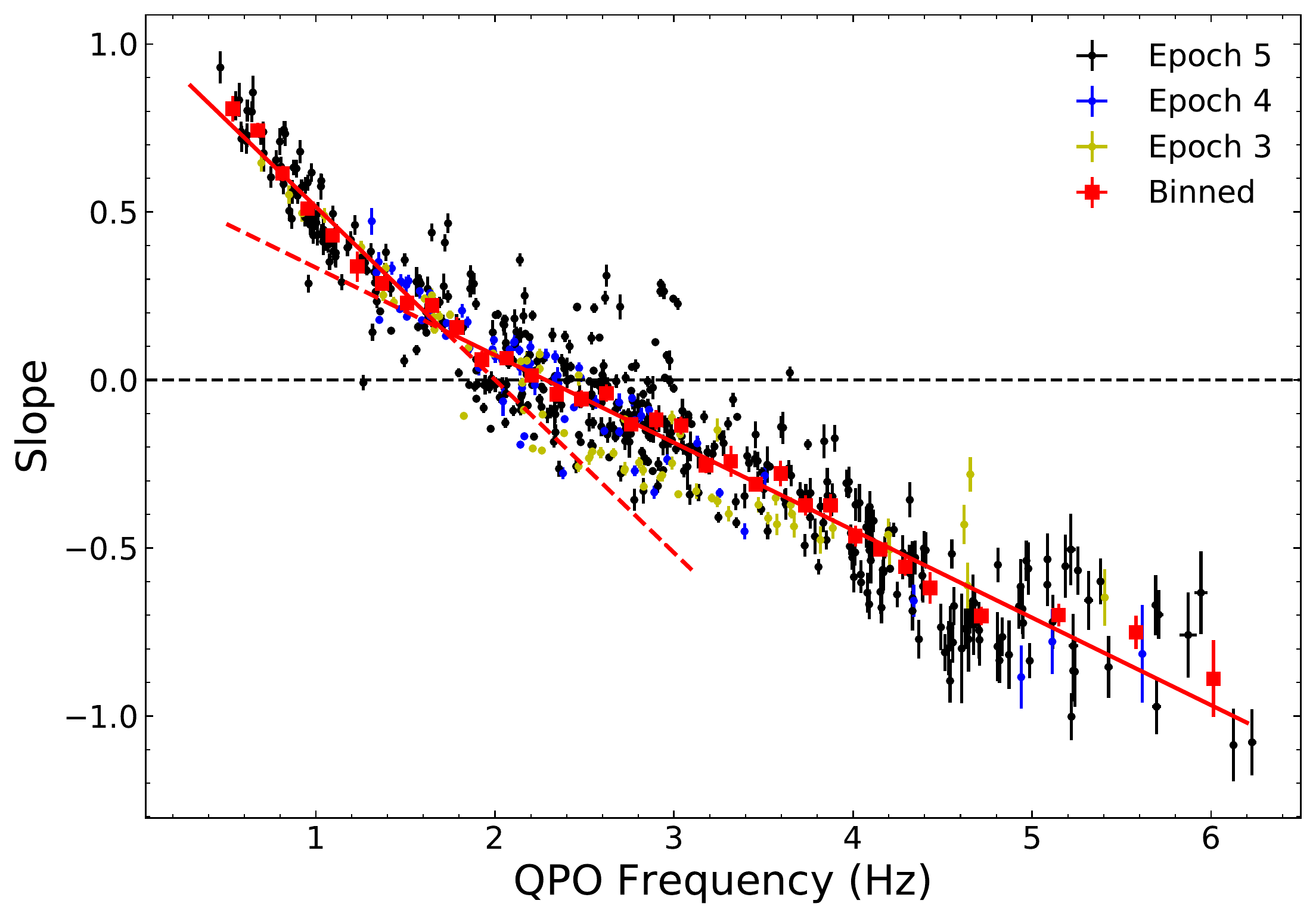}}}
\caption{Slope of the QPO lag-energy spectra for the fundamental of the type-C QPO in GRS 1915+105. The solid lines indicate the best-fitting broken line to the binned data. The dashed lines are the continuation of the broken line, drawn for comparison with the data in those parts of the plot. }
\label{fig:phase_fit}
\end{figure}

\begin{figure}
\centering
\resizebox{\columnwidth}{!}{\rotatebox{0}{\includegraphics{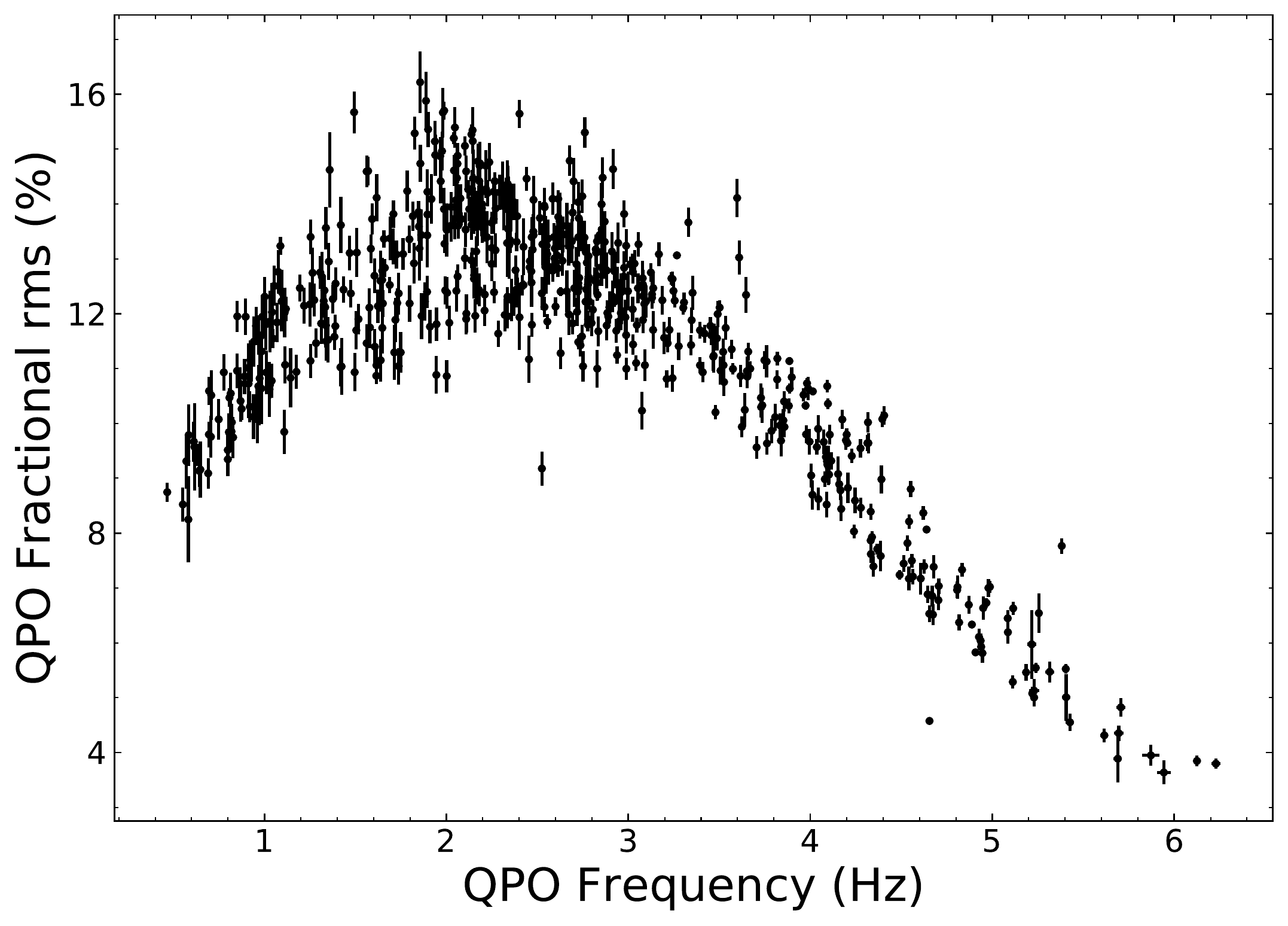}}}
\caption{QPO fractional rms in the full PCA band as a function of the QPO frequency for the fundamental of the type-C QPO in GRS 1915+105.}
\label{fig:fre_rms}
\end{figure}

\begin{figure*}
\centering
\resizebox{2\columnwidth}{!}{\rotatebox{0}{\includegraphics{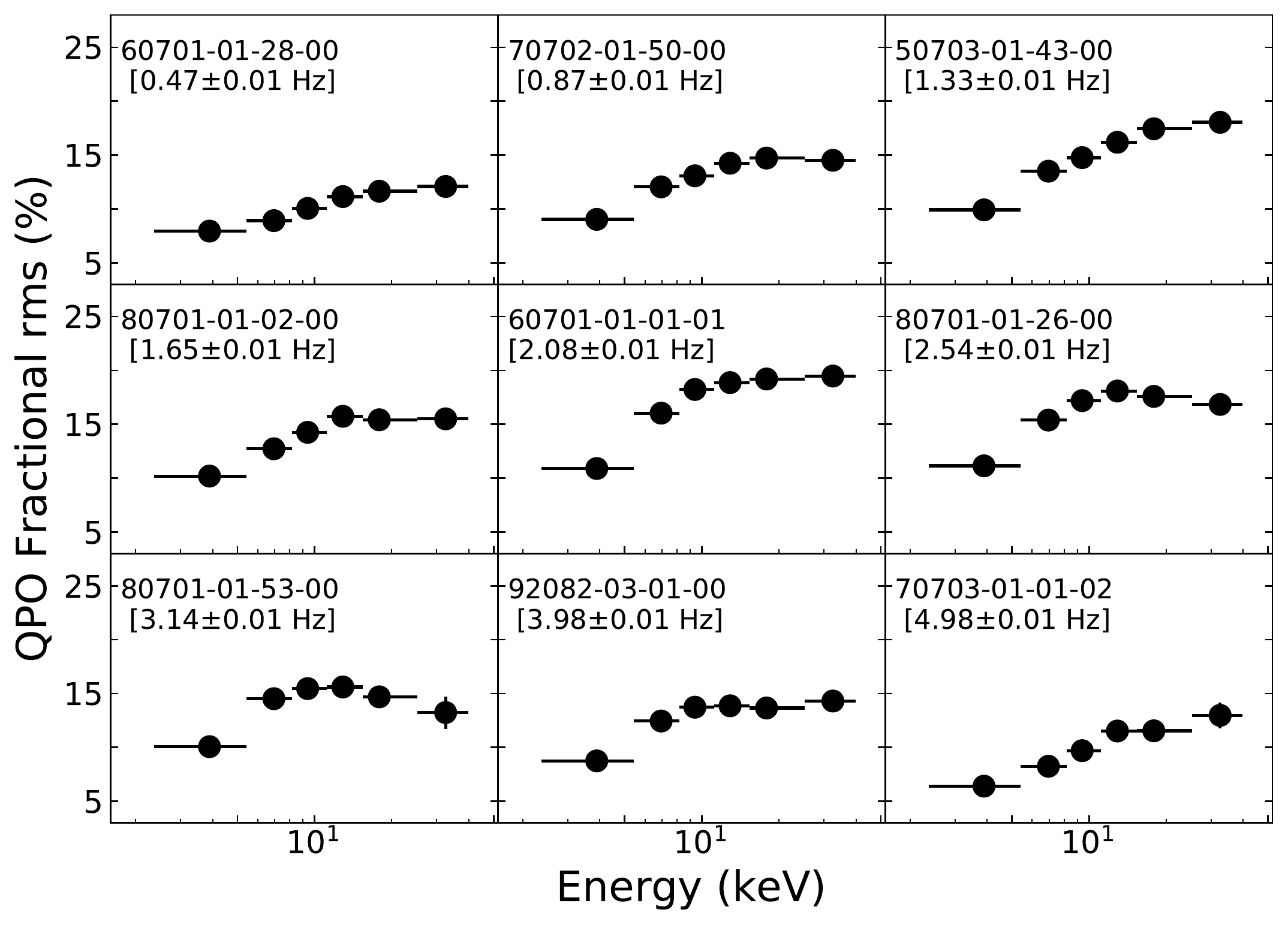}}}
\caption{Representative examples of the QPO rms spectra for the fundamental of the type-C QPO in GRS 1915+105. The observations we show here are the same ones as in Fig. \ref{fig:pds}. Observation ID and QPO frequency are listed in each panel.}
\label{fig:rms_spectra}
\end{figure*}

\begin{figure}
\centering
\resizebox{\columnwidth}{!}{\rotatebox{0}{\includegraphics{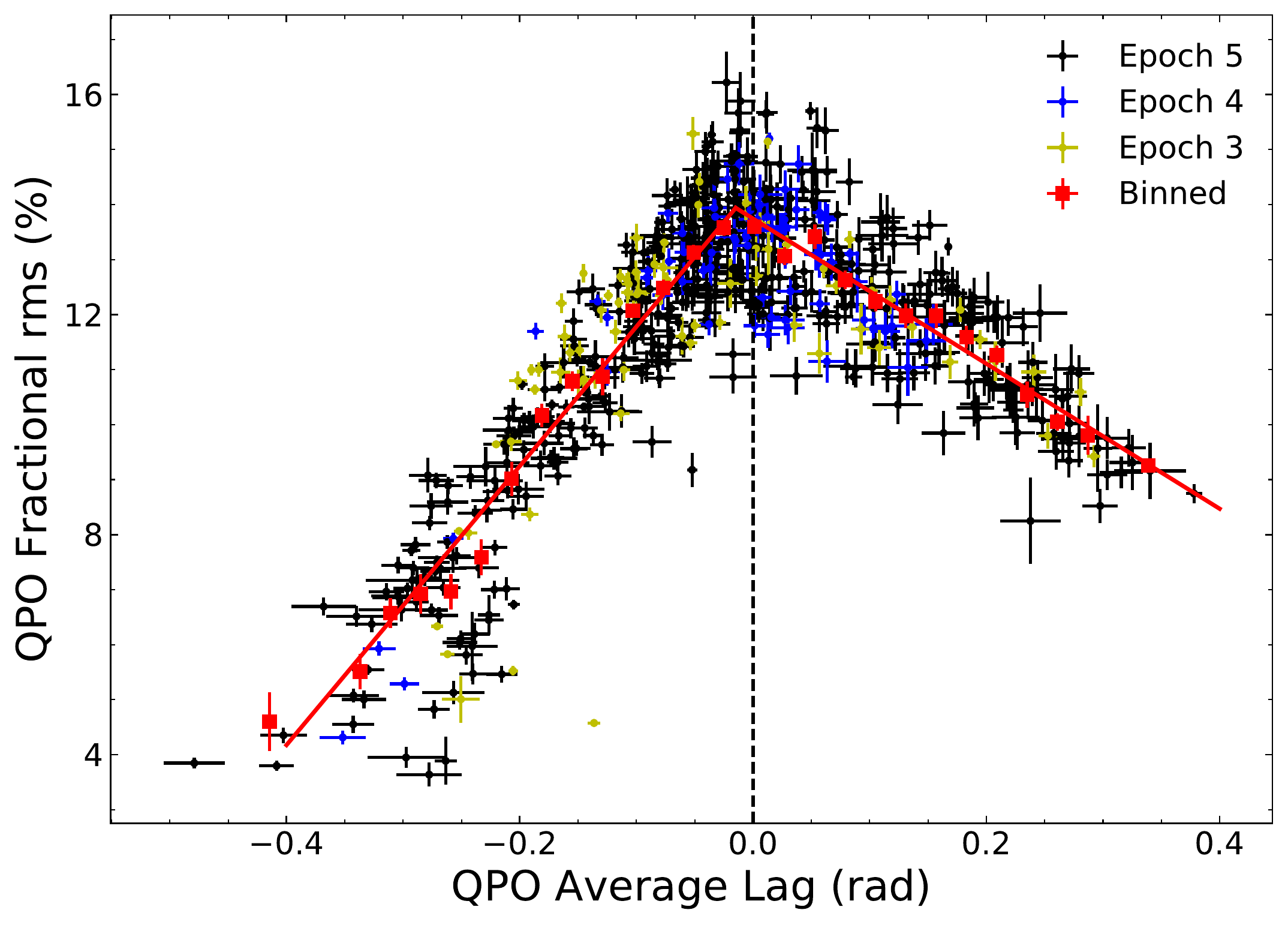}}}
\caption{QPO fractional rms in the full PCA band as a function of the average phase lags at the fundamental of the type-C QPO in GRS 1915+105. The phase lags were calculated in the same bands as in Fig. \ref{fig:pds}. The solid lines are the best-fitting broken line to the binned data.}
\label{fig:phase_rms}
\end{figure}

Fig. \ref{fig:pds} shows 9 representative examples of power and lag-frequency spectra for observations with different QPO frequencies. The power spectra were calculated for the full PCA band, while the phase lags were computed between the $2-5.7$ keV and $5.7-15$ keV bands\footnote{Throughout this paper, the phase lags between the $2-5.7$ keV and $5.7-15$ keV bands will be referred to as ``average phase lags''.}. In all cases a strong (up to $\sim16\%$ fractional rms), narrow ($Q \sim 4-37$) QPO is present in the PDS, together with a band-limited noise component.
In Tab. \ref{tab:QPO property} we summarise the main properties of those QPOs in our sample. In Fig. \ref{fig:hist} we show the histogram of the main properties of those QPOs. The QPO frequency ranges from $\sim0.4$ Hz to $\sim6.3$ Hz with a peak in the distribution at $\sim2.2$ Hz. The majority of the QPOs have a $Q$ factor of $\sim5-15$, with a peak of the distribution at $Q\sim10$, and in some cases the $Q$ factor of the QPO can reach up to $Q\sim30$. The fractional rms amplitude of the QPO in the full PCA band ranges from $\sim3\%$ to $\sim17\%$, with a peak of the distribution at $\sim12\%$. The average phase lags at the QPO fundamental between 2--5.7 keV and 5.7--15 keV bands ranges from $\sim-0.4$ rad to $\sim0.4$ rad, with a peak of the distribution at $\sim0$ rad. In 471 cases the QPO shows a clear second harmonic and in 141 cases the QPO shows a subharmonic. In all cases, if a subharmonic is present in the PDS, a second-harmonic is always observed. The second-harmonic and subharmonic frequencies are consistent with being two times and half that of the QPO fundamental, respectively. Their fractional amplitudes are significantly smaller than that of the QPO fundamental.

From the lag-frequency spectra, we find clear difference between QPOs below and above $\sim 2$ Hz. When the QPO frequency is above $\sim 2$ Hz, a dip-like feature centered on the QPO frequency is always present in the lag-frequency spectra. In addition, if a second harmonic is present, a peak-like feature is visible as well near the second-harmonic frequency. However, when the QPO frequency is below $\sim 2$ Hz, no such features are apparent, and the lag-frequency spectra are roughly flat at low frequencies.


\subsection{QPO Fundamental Lags}\label{sec:fundamental}

In Fig. \ref{fig:fre_phase}, we plot the average phase lags at the fundamental of the type-C QPO (hereafter QPO average lags) in GRS 1915+105 as a function of the centroid frequency of the QPO (hereafter QPO frequency). As previously shown by \citet{Reig2000}, \citet{Qu2010} and \citet{Pahari2013}, it is apparent in the figure that the QPO average lags decrease with QPO frequency, changing sign from positive to negative at a QPO frequency of around 2 Hz. In addition, we notice that the slope of the correlation is different when the QPO frequency is below or above $\sim 2$ Hz. The break was already seen in the Fig. 3 of \citet{Reig2000}, although they did not mention it in their paper. To check whether there is a significant change in the slope, we rebinned the data in Fig. \ref{fig:fre_phase} and fitted them both with a broken line and a straight line. We compared the fits using an F-test, and find that the broken-line fit is better than the straight-line fit at a confidence level of $>5\sigma$. The break obtained from the broken-line fit is at $\nu_{\rm QPO}=1.80\pm0.11$ Hz. The slope below the break is $-0.21\pm0.02$, and above the break is $-0.10\pm0.01$. The frequencies at which the QPO average lags reach zero are $1.99 \pm 0.08$ Hz and $2.20 \pm 0.18$ Hz for the lines below and above the break, respectively. It is important to mention that the scatter above $\sim 2$ Hz is significantly larger than that below $\sim 2$ Hz, and it is larger than the statistical fluctuation expected from the error bars.

In Fig. \ref{fig:phase_spectra}, we show representative QPO lag-energy spectra for observations with different QPO frequencies. We find that for the QPOs with centroid frequencies below $\sim 2$ Hz, the QPO lags are always positive and increase monotonically with energy, for the QPOs with centroid frequencies around 2 Hz, the QPO lags are nearly zero in all energy bands, while for the QPOs with centroid frequencies above $\sim 2$ Hz, the QPO lags are always negative and decrease monotonically with energy. \citet{Pahari2013} and \citet{Eijnden2017} found broad features in the QPO lag-energy spectra of several observations around an energy of $6-7$ keV, which might be related to reflection. Because here we used a relatively low energy resolution for the lag-energy spectra, such features are not obvious in our results. 

For each observation, we fitted the lag-energy spectra with the function $\Delta\phi(E)=\alpha {\rm log} E/E_{\rm ref}$, where $E_{\rm ref}$ is the energy of the reference band. As shown in Fig. \ref{fig:phase_fit}, the slope of the lag-energy spectra, $\alpha$, shows a similar evolution with QPO frequency as the QPO average lags. The slope also decreases with QPO frequency, and changes sign from positive to negative at around 2 Hz. In addition, we also observe a clear break in this relation at $\sim 2$ Hz. Again, we fitted the data in Fig. \ref{fig:phase_fit} both with a broken and a straight line. An F-test indicates that the broken-line fit is better than the straight-line fit at a confidence level of $>5\sigma$. The break obtained from the best fit is at $\nu_{\rm QPO}=1.72\pm0.10$ Hz. The slope below the break is $-0.52\pm0.03$, and above the break is $-0.26\pm0.01$.

In Fig. \ref{fig:fre_rms} we show the QPO fractional rms in the full PCA band as a function of the QPO frequency. The evolution of the QPO fractional rms is similar to that reported by \citet{Yan2013}: the QPO fractional rms first increases with QPO frequency and then decreases, reaching its maximum at a QPO frequency of $\sim2$ Hz. In Fig. \ref{fig:rms_spectra} we show several examples of the QPO rms spectra. In all cases the QPO rms spectra are hard: the QPO rms first increases at low energies, and then turns to flat or slightly decreases above $\sim 12$ keV. Such kind of QPO rms spectra has been observed in many other BHBs, such as XTE J1550--564~\citep{Rodriguez2004}, GS 1124--68~\citep{Belloni1997}, GX 339--4~\citep{Zhang2017}, and \citep{Huang2018}.

From Fig. \ref{fig:fre_phase} and \ref{fig:fre_rms}, it is apparent that the behaviours of both the QPO average lags and the QPO fractional rms change significantly at $\sim2$ Hz. In Fig. \ref{fig:phase_rms}, we plot the QPO fractional rms as a function of the QPO average lags. As one can see from this figure, the QPO fractional rms first increases as the QPO average lags go from negative to zero, reaching its maximum at around zero lag, and then the QPO fractional rms decreases as the QPO average lags increase further. We rebinned the data and fitted them with a broken line. The break obtained from the fit is at $\Delta\phi=0.015\pm0.008$ rad. This confirms our previous result that the behaviour of the QPO average lags changes at around zero lag.

\begin{figure}
\centering
\resizebox{\columnwidth}{!}{\rotatebox{0}{\includegraphics{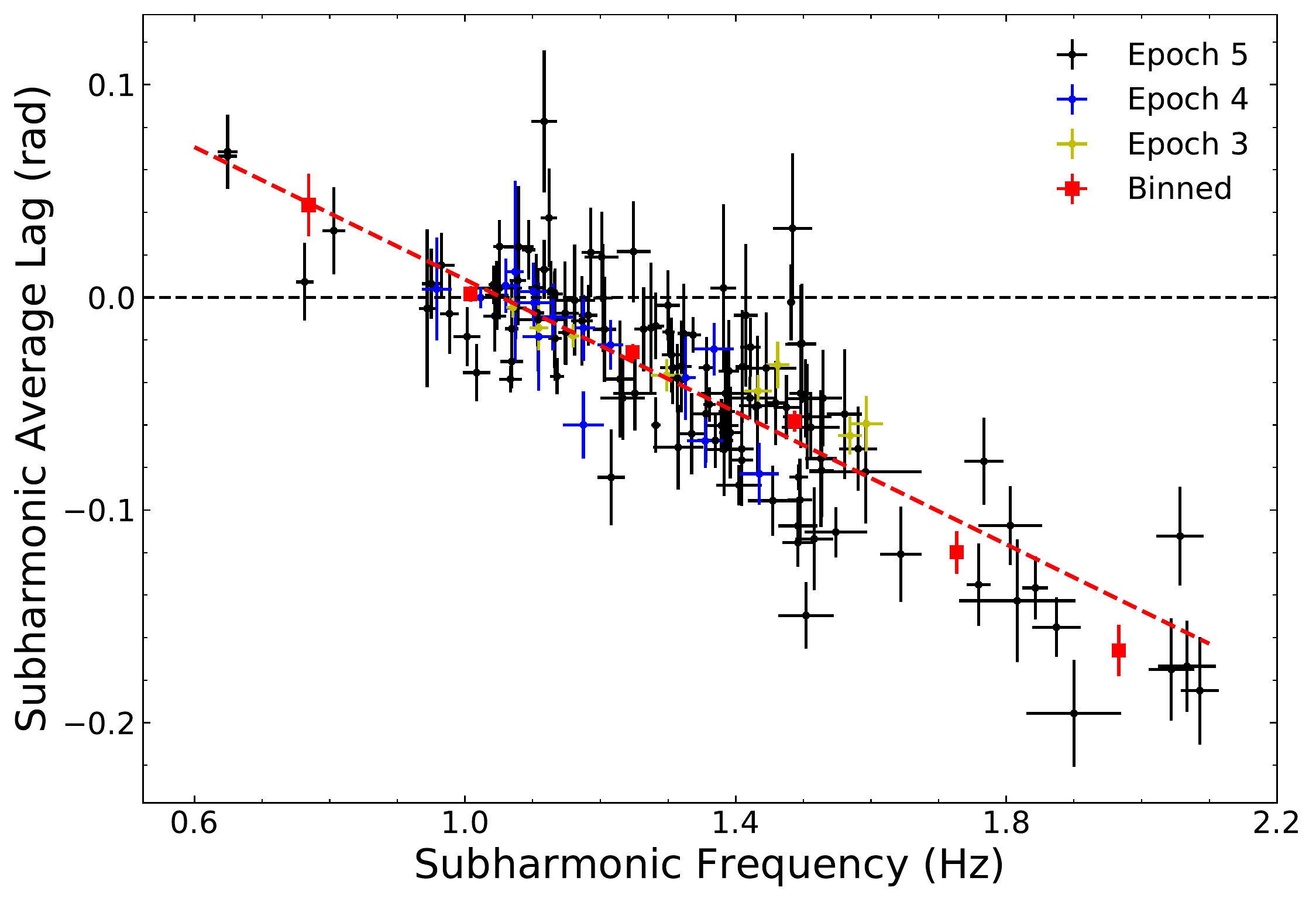}}}
\caption{Average phase lags at the subharmonic frequency of the type-C QPO in \target as a function of the subharmonic frequency. The phase lags were calculated in the same bands as in Fig. \ref{fig:pds}. The dashed line is the best-fitting straight line to the binned data.}
\label{fig:fre_phase_sub}
\end{figure}

\begin{figure}
\centering
\resizebox{\columnwidth}{!}{\rotatebox{0}{\includegraphics{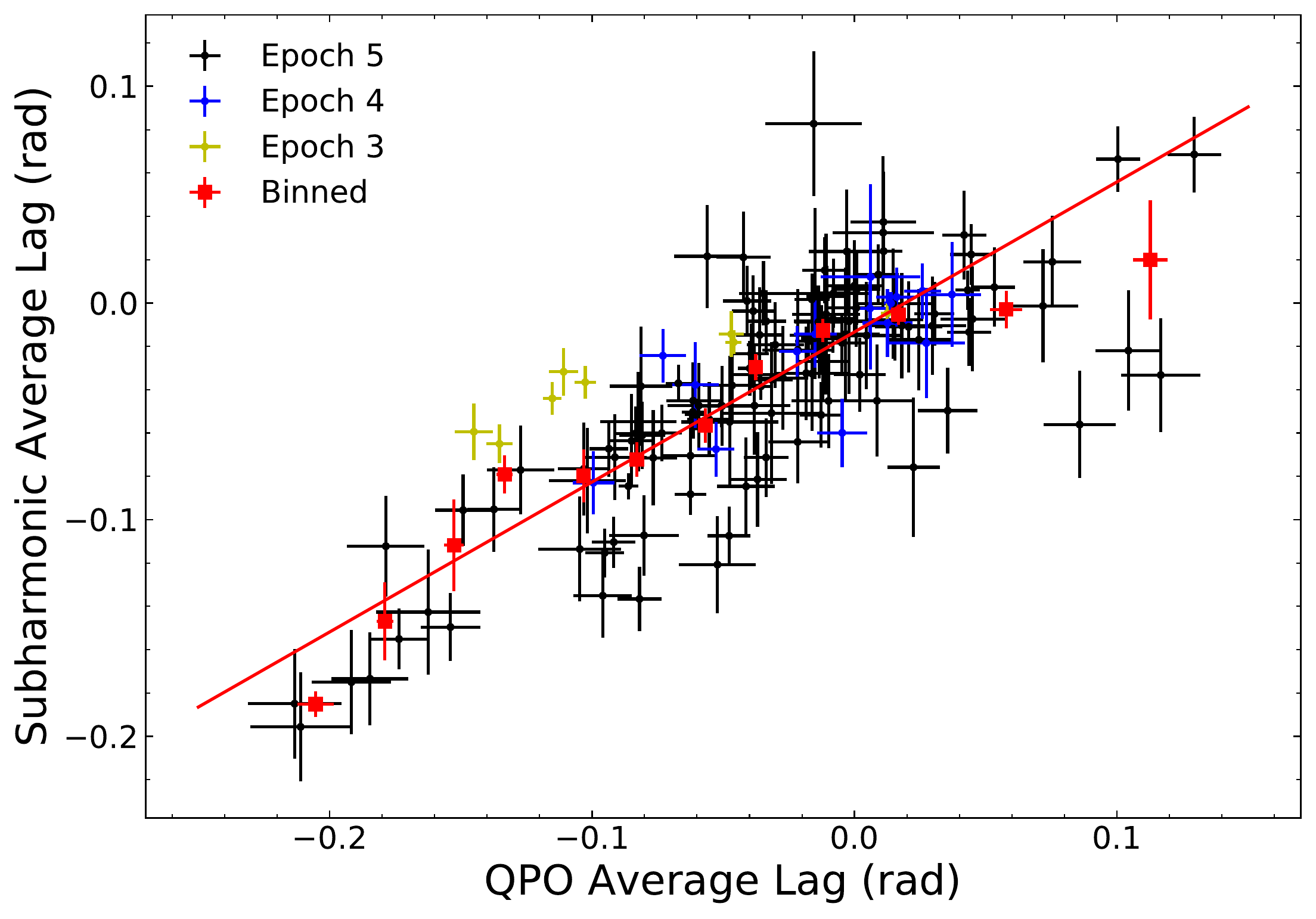}}}
\caption{Subharmonic average lags as a function of the QPO average lags for the type-C QPOs with a subharmonic in GRS 1915+105. The phase lags were calculated in the same bands as in Fig. \ref{fig:pds}. The dashed line is the best-fitting straight line to the binned data.}
\label{fig:fund_sub_lag}
\end{figure}

\subsection{Subharmonic Lags}\label{sec:subharmonic}

\begin{figure}
\centering
\resizebox{\columnwidth}{!}{\rotatebox{0}{\includegraphics{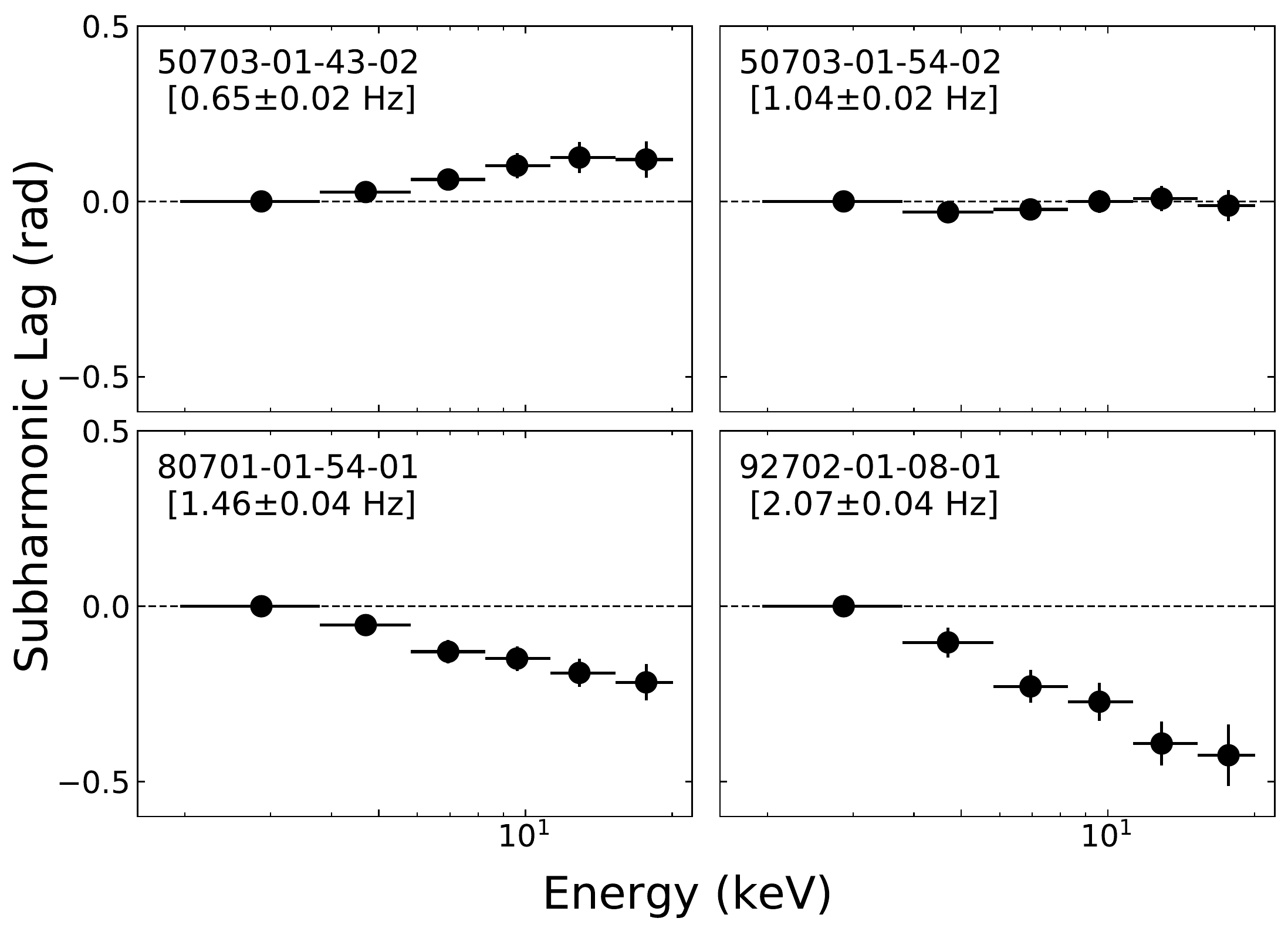}}}
\caption{Examples of the lag-energy spectra for the subharmonic of the type-C QPO in GRS 1915+105. Observation ID and subharmonic frequency are listed in each panel.}
\label{fig:phase_spectra_sub}
\end{figure}

\begin{figure}
\centering
\resizebox{\columnwidth}{!}{\rotatebox{0}{\includegraphics{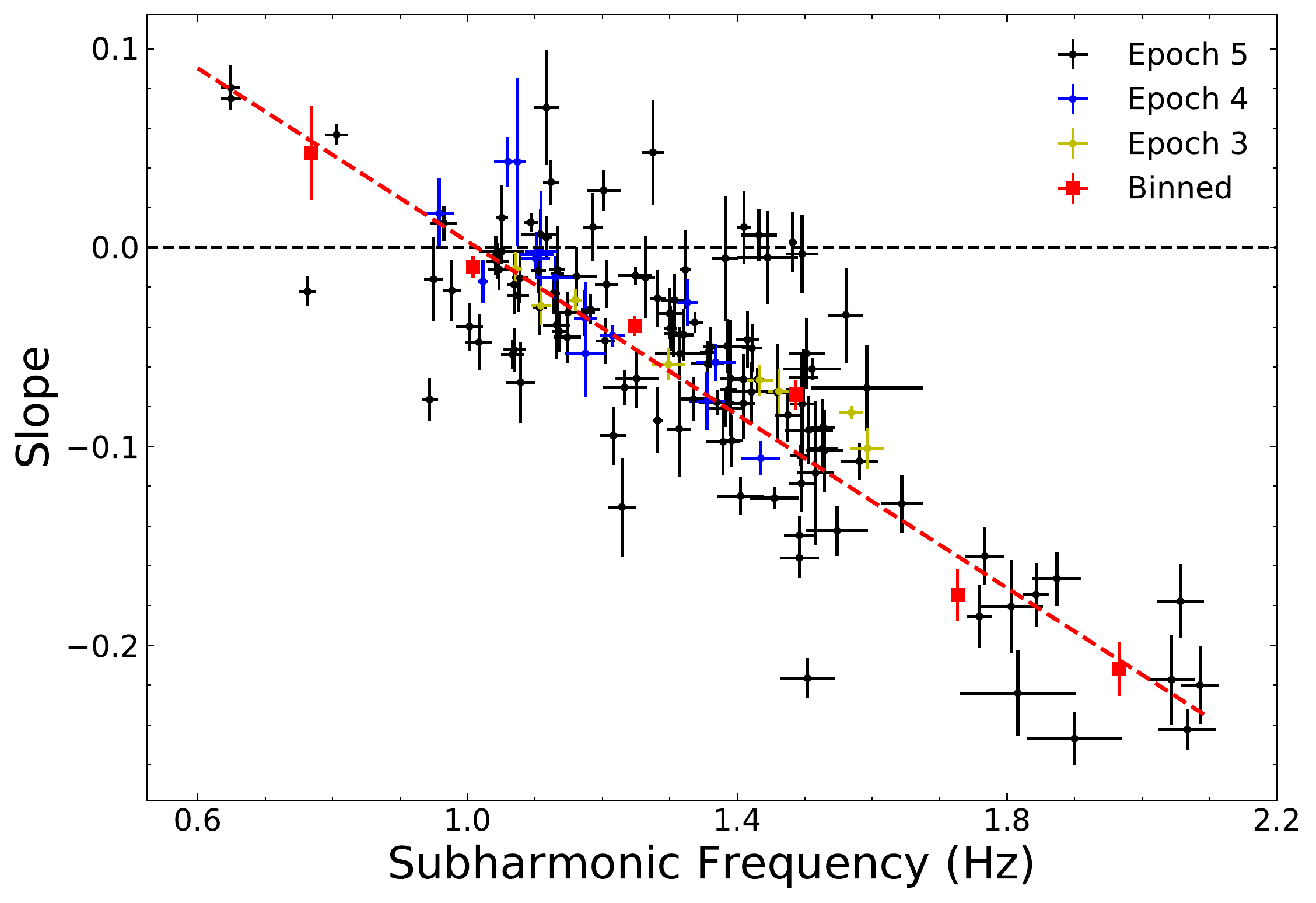}}}
\caption{Slope of the lag-energy spectra for the subharmonic of the type-C QPO in GRS 1915+105. The dashed line is the best-fitting straight line to the binned data.}
\label{fig:phase_fit_sub}
\end{figure}

\begin{figure}
\centering
\resizebox{\columnwidth}{!}{\rotatebox{0}{\includegraphics{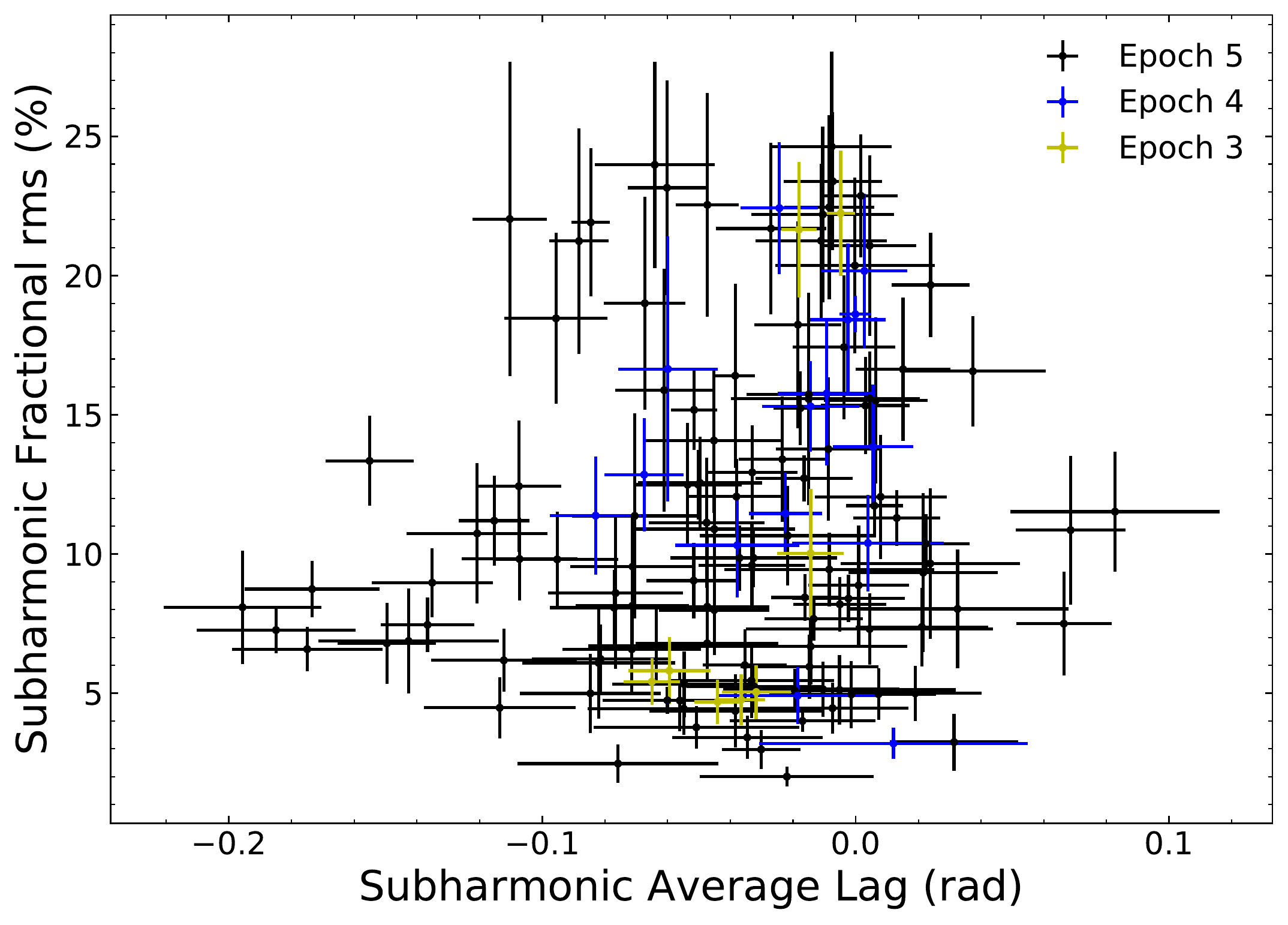}}}
\caption{Fractional rms in the full PCA band as a function of the average phase lags for the subharmonic of the type-C QPO in GRS 1915+105.}
\label{fig:phase_rms_sub}
\end{figure}

In Fig. \ref{fig:fre_phase_sub}, we show the average phase lags at the frequency of the subharmonic of the type-C QPO (hereafter subharmonic average lags) in GRS 1915+105 as a function of the centroid frequency of the subharmonic of the QPO (hereafter subharmonic frequency). The behaviour of the subharmonic average lags is very similar to that of the QPO average lags. The subharmonic average lags decrease with subharmonic frequency, changing sign from positive to negative at a subharmonic frequency of $\sim 1$ Hz. It is not clear whether there is a break in this relation due to the sparsity of data below 1 Hz. We then rebinned the data in Fig. \ref{fig:fre_phase_sub} and fitted them with a straight line. The slope obtained from the fit is $-0.16\pm0.02$, which is in between the slope below and above the break in Fig. \ref{fig:fre_phase}. The frequency at which the subharmonic average lags reach zero is $1.05 \pm 0.19$ Hz obtained from the fit, which is consistent with being half of the frequency at which the QPO average lags change sign within error.

In Fig. \ref{fig:fund_sub_lag}, we show the subharmonic average lags as a function of the QPO average lags for the type-C QPOs with a subharmonic. As we can expect from Fig. \ref{fig:fre_phase} and \ref{fig:fre_phase_sub}, the subharmonic average lags generally increase with the QPO average lags. A possible break seems to exist at QPO average lags $\sim 0$ rad in Fig. \ref{fig:fund_sub_lag}. To test this we rebinned the data and fitted them both with a broken and a straight line. We found that the broken-line fit is marginally better than the straight-line fit, at a confidence level of $< 3 \sigma$. The slope ($0.69 \pm 0.07$) and intercept ($-0.013 \pm 0.006$) obtained from the straight-line fit are not consistent with being 0.5 and 0 within errors. This means that the amplitude of the subharmonic average lags is not exactly half that of the QPO average lags.

In Fig. \ref{fig:phase_spectra_sub} we show four representative examples of the lag-energy spectra for the subharmonic. We find that, similar to the QPO fundamental, when the subharmonic frequency is below $\sim 1$ Hz, the subharmonic lags slightly increase with energy, when the subharmonic frequency is around $\sim 1$ Hz, the subharmonic lags are nearly zero in all energy bands, while when the subharmonic frequency is above $\sim 1$ Hz, the subharmonic lags decrease monotonically with energy. We fitted the lag-energy spectra of the subharmonic with the same function that we used for the QPO fundamental. As is apparent in Fig. \ref{fig:phase_fit_sub}, the slope also decreases with subharmonic frequency, and changes sign at a subharmonic frequency of $\sim 1$ Hz. The slope obtained from the fit is $-0.22\pm0.02$.

In Fig. \ref{fig:phase_rms_sub} we show the fractional rms in the full PCA band of the subharmonic of the QPO (hereafter subharmonic fractional rms) as a function of the subharmonic average lags. No clear correlation is apparent as the one we found in the QPO fractional rms and QPO average lags relation (Fig. \ref{fig:phase_rms}).

\begin{figure}
\centering
\resizebox{\columnwidth}{!}{\rotatebox{0}{\includegraphics{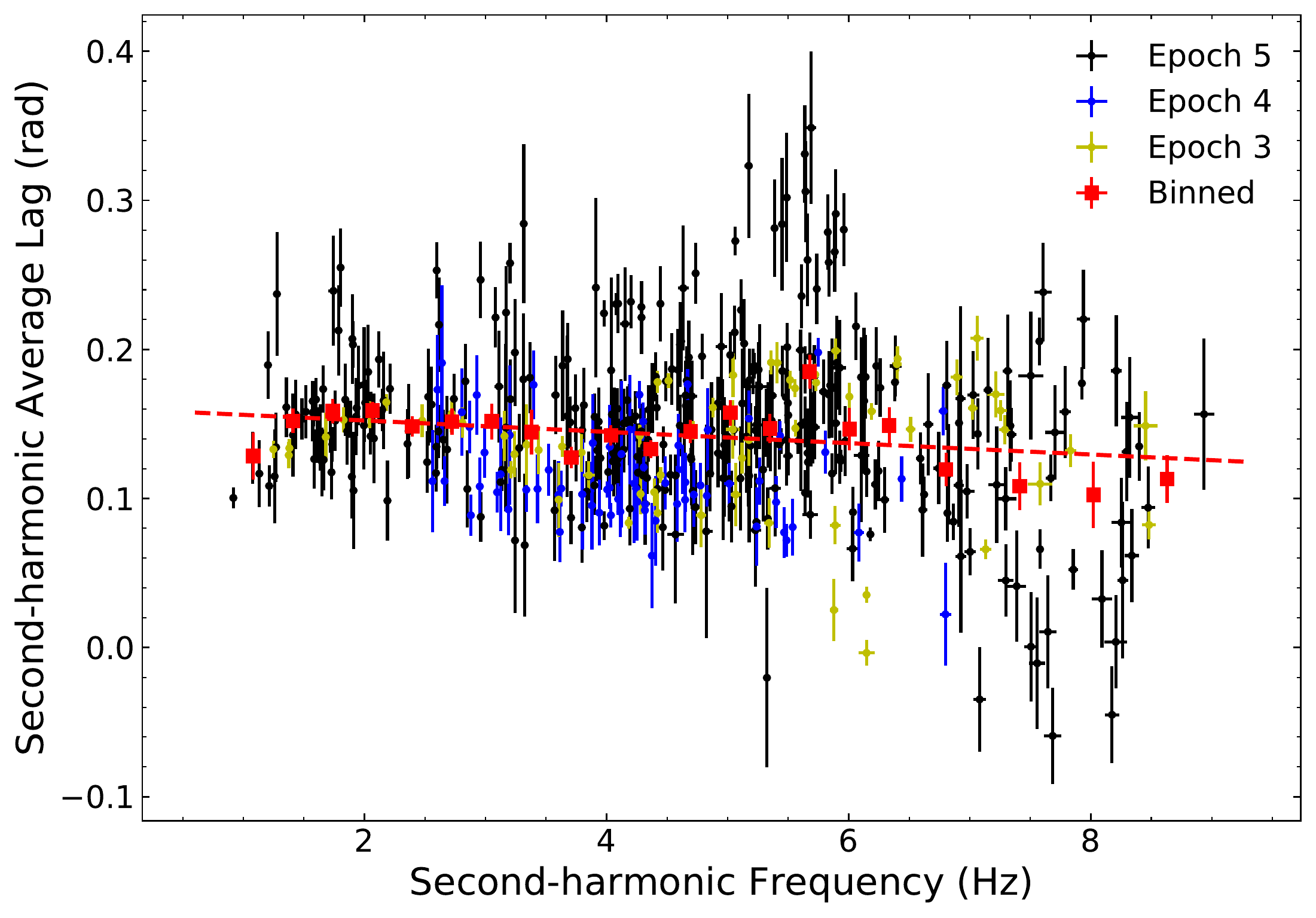}}}
\caption{Average phase lags at the second-harmonic frequency of the type-C QPO in \target as a function of the second-harmonic frequency. The phase lags were calculated in the same bands as in Fig. \ref{fig:pds}. The dashed line is the best-fitting straight line to the binned data.}
\label{fig:fre_phase_second}
\end{figure}

\begin{figure}
\centering
\resizebox{\columnwidth}{!}{\rotatebox{0}{\includegraphics{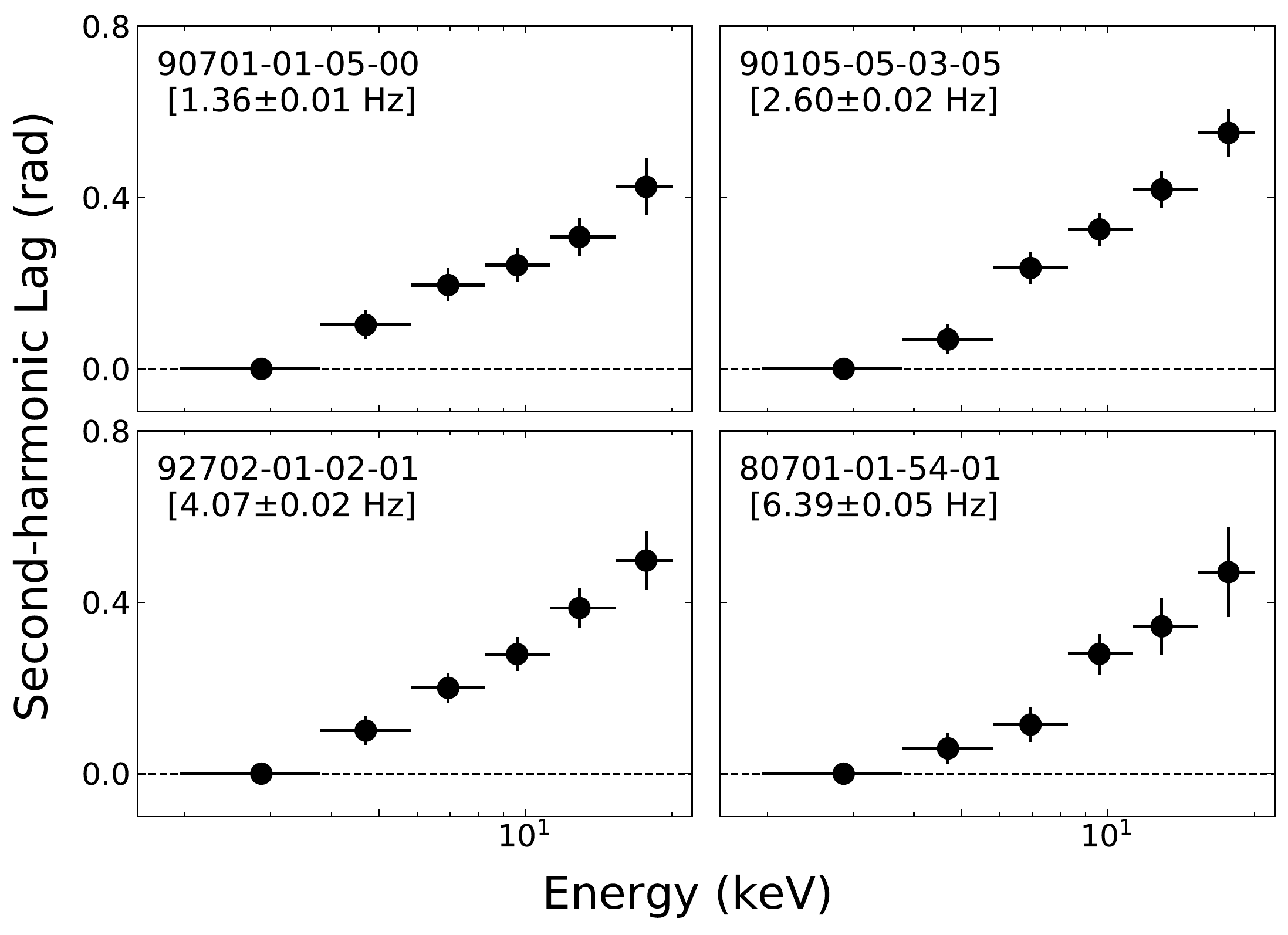}}}
\caption{Examples of the lag-energy spectra for the second harmonic of the type-C QPO in GRS 1915+105. Observation ID and second-harmonic frequency are listed in each panel.}
\label{fig:phase_spectra_second}
\end{figure}

\begin{figure}
\centering
\resizebox{\columnwidth}{!}{\rotatebox{0}{\includegraphics{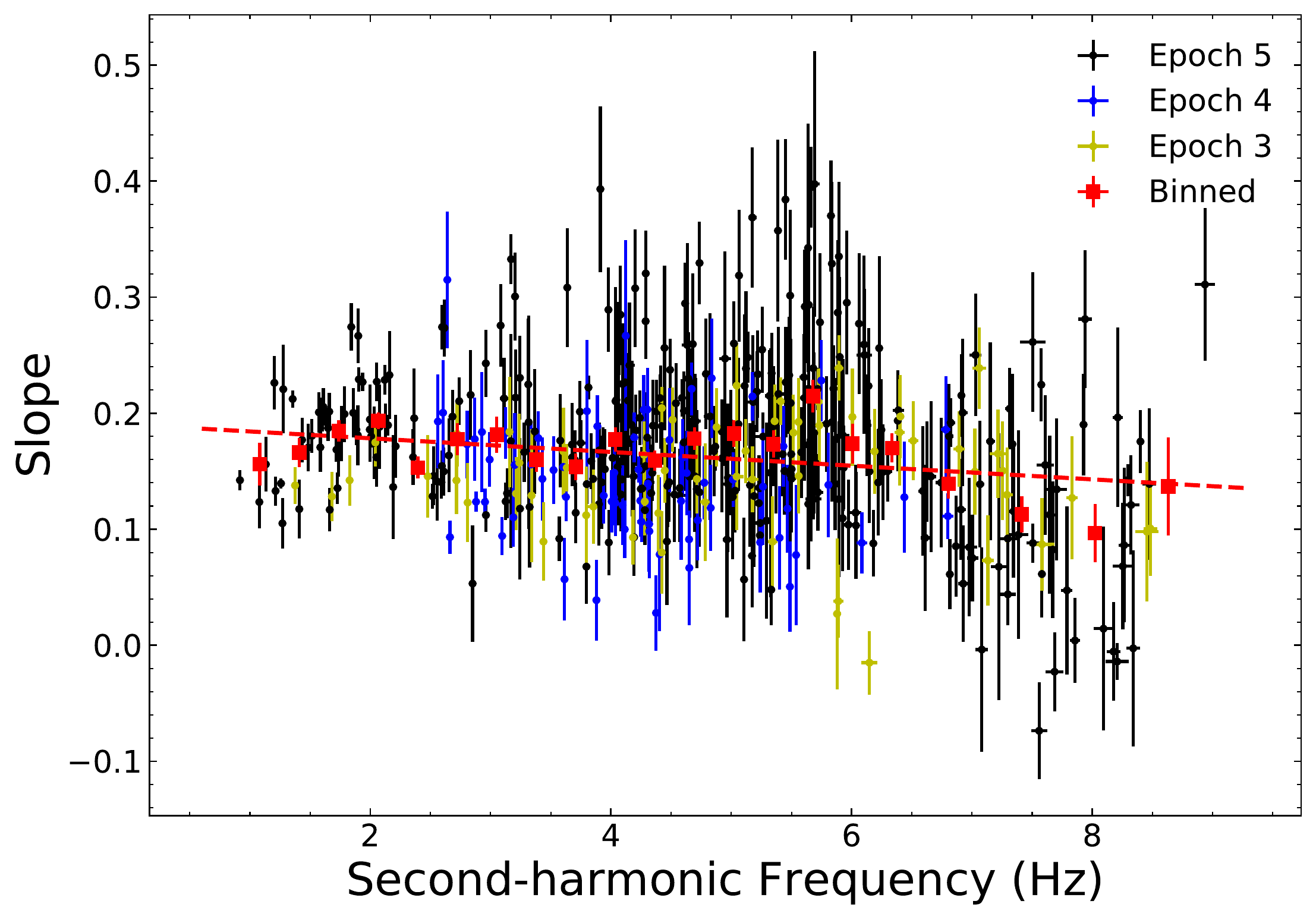}}}
\caption{Slope of the lag-energy spectra for the second harmonic of the type-C QPO in GRS 1915+105. The dashed line is the best-fitting straight line to the binned data.}
\label{fig:phase_fit_second}
\end{figure}

\subsection{Second-harmonic Lags}\label{sec:second_harmonic}

In Fig. \ref{fig:fre_phase_second} we show the average phase lags at the frequency of the second harmonic of the type-C QPO (hereafter second-harmonic average lags) in GRS 1915+105 as a function of the centroid frequency of the second harmonic of the QPO (hereafter second-harmonic frequency). Compared to the QPO fundamental and subharmonic, the phase-lag behaviour at the frequency of the second harmonic is quite different: almost all the second harmonics show a positive lag, and the second-harmonic average lags remain more or less constant at $\Delta\phi\sim0.15$ rad. No changing sign from positive to negative is observed. We rebinned the data in Fig. \ref{fig:fre_phase_second} and fitted them with a straight line. The slope obtained from the fit is $-0.004\pm0.002$. Note that the scatter of the second-harmonic lags is very large, especially at high frequencies. 

In Fig. \ref{fig:phase_spectra_second} we show four examples of the lag-energy spectra for the second harmonic. We find that the second-harmonic lags always increase monotonically with energy. We fitted the lag-energy spectra of the second harmonic with the same function that we used for the QPO fundamental and subharmonic. The slope as a function of the second-harmonic frequency is shown in Fig. \ref{fig:phase_fit_second}. Although the scatter of the slope is very large, it is apparent that the evolution of the slope with second-harmonic frequency is similar to that of the second-harmonic average lags. We also fitted the rebinned data in Fig. \ref{fig:phase_fit_second} with a straight line, and the slope obtained from the fit is $-0.006\pm0.002$.

\begin{figure}
\centering
\resizebox{\columnwidth}{!}{\rotatebox{0}{\includegraphics{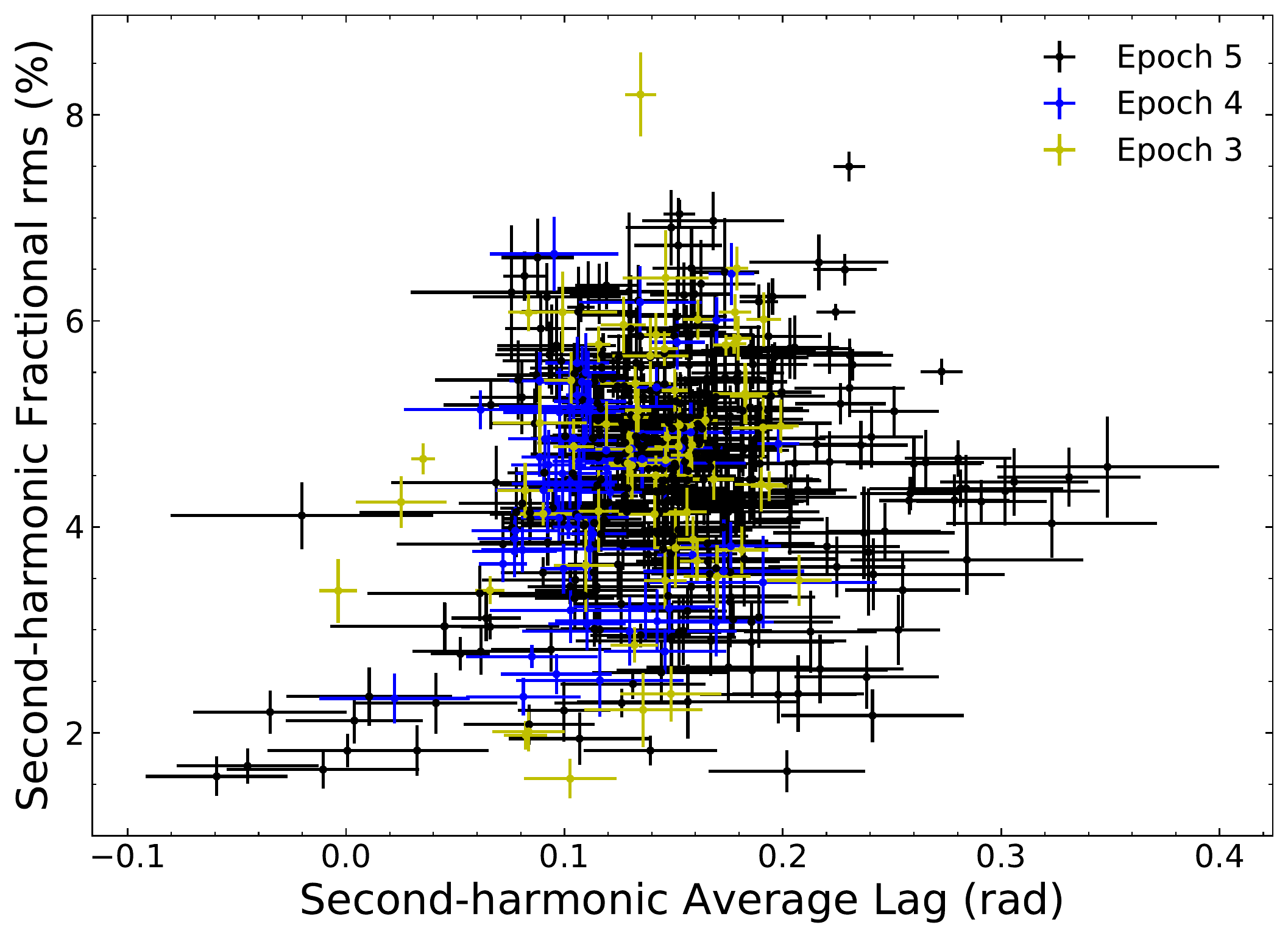}}}
\caption{Fractional rms in the full PCA band as a function of the average phase lags for the second harmonic of the type-C QPO in GRS 1915+105.}
\label{fig:phase_rms_second}
\end{figure}

In Fig. \ref{fig:phase_rms_second} we show the fractional rms in the full PCA band of the second harmonic of the QPO (hereafter second-harmonic fractional rms) as a function of the second-harmonic average lags. Similar to the subharmonic, no clear correlation is found in the second-harmonic fractional rms and second-harmonic average lags relation.

\section{Discussion}\label{sec:Discussion}

We present the first systematic analysis of the phase lags associated with the type-C QPOs in GRS 1915+105, based on the largest QPO sample of this source to date. Our sample comprises of 620 {\it RXTE} observations with type-C QPO, with a frequency ranging from 0.4 Hz to 6.3 Hz. We found that: 
\begin{enumerate}
\item the relation between the QPO fractional rms and the QPO average lags can be well fitted with a broken line (Fig. \ref{fig:phase_rms});
\item the phase-lag behaviour of the subhamonic of the QPO is quite similar to that of the QPO fundamental; on the contrary, the second harmonic of the QPO show a very different phase-lag behaviour;
\item for both the QPO fundamental and its (sub)harmonics, the slope of the lag-energy spectra shows a similar evolution with frequency as the average phase lags;
\item we confirm that the QPO average lags decrease with QPO frequency, and change sign from positive to negative at around 2 Hz \citep{Reig2000,Qu2010,Pahari2013}. 
\item we find that the slope of the relation between the QPO average lags and the QPO frequency (Fig. \ref{fig:fre_phase}) is significantly different when the QPO frequency is below or above $\sim2$ Hz;
\end{enumerate}
We discuss each of these findings below.

\subsection{QPO Fundamental lags}

Based on a systematic analysis of 15 Galactic BHBs, \citet{Eijnden2017} found that, at high QPO frequencies ($>2$ Hz) high-inclination sources generally show soft QPO lags, while low-inclination sources show hard QPO lags. As a high-inclination source \citep{Mirabel1994}, \target is consistent with other high-inclination sources that exhibit soft QPO lags when the QPO frequency is above 2 Hz. However, at low fundamental frequencies ($<2$ Hz) all other sources tend to show zero QPO lags, whereas \target shows hard QPO lags that increase in magnitude towards low QPO frequencies. This unique QPO lag behaviour, especially the change in lag sign, makes \target an interesting source to study the origin of the QPO lags. In our analysis, we noticed a significant break in the relation between the QPO average lags and the QPO frequency (Fig. \ref{fig:fre_phase}). The break frequency is consistent with the frequency at which the QPO average lags change sign. This supports scenarios in which something related to the accretion process changes as the lags change sign. Presently, the most promising model for type-C QPOs is the Lense-Thirring precession of the entire hot inner flow proposed by \citet{Ingram2009}. In this model, the outer radius of the inner flow, $r_{0}$, is set by the truncation radius of the accretion disk; the inner radius, $r_{\rm i}$, is set by the radius of the warps due to the misalignment between black hole spin and inner flow. \citet{Miller2013} estimated the black hole spin as $a_{*}=0.98\pm0.01$, and \citet{Reid2014} measured the mass of the black hole of $12.4_{-1.8}^{+2.0}~M_{\sun}$ for GRS 1915+105. Taking these values, we estimated $r_{0} \sim 16~R_{\rm g}$ and $r_{\rm i} \sim 10~R_{\rm g}$ for a 2 Hz QPO\footnote{Same as described in \citet{Ingram2009}, here we assumed that the scale-height $h/r=0.2$ and the index of the surface density $\zeta = 0$ in the calculation.} according to equation (2) and (3) of \citet{Ingram2009}.

Several models have been proposed to explain the change in lag sign. For instance, \citet{Nobili2000} presented a thermal Comptonisation model based on the results of \citet{Reig2000}. Their model requires a two-component corona: an inner hot and optically thick component and an outer cooler and optically thinner component. The corona up-scatters seed photons from an accretion disc that is truncated at an inner radius $r_{\rm in}$. To explain the dependence of the QPO lags upon QPO frequency, \citet{Nobili2000} assume that at high QPO frequencies ($>2$ Hz) the disc is truncated at small radii ($\sim6$ Rg) while the inner corona is so optically thick ($\tau\geq100$) that, with a moderate temperature of $\sim15$ keV, the corona is able to up-scatter all the soft disc photons up to $\sim15$ keV. The hard photons will then suffer down-scatterings in the outer cooler and optically thinner component that will eventually result in a soft lag. On the contrary, at lower QPO frequencies ($<2$ Hz), the disc is assumed to be truncated at larger radii, and the inner corona is assumed to become optically thinner, thus the soft photons from the disc will only experience Compton up-scattering and the lags will be hard. In this model, the change in lag sign is due to the change of the inner disc radius and the optical depth of the corona.

\target is one of the sources showing QPO frequency shifts, $\Delta\nu_{0}$, between energy bands. These shifts reach up to $\sim0.5$ Hz when the QPO is at 5 Hz, comparing the QPO frequency between the $2-5$ keV and $13-18$ keV bands \citep{Qu2010}. The energy dependence of the QPO frequency in \target has been studied by \citet{Qu2010} and \citet{Yan2012}. They found that when the QPO frequency is below 2 Hz, the QPO frequency decreases with energy (negative $\Delta\nu_{0}$) whereas, when the QPO frequency is above 2 Hz, the QPO frequency increases with energy (positive $\Delta\nu_{0}$). Interestingly, the frequency where $\Delta\nu_{0}$ changes sign is consistent with that at which the QPO average lags also change sign. To explain the change in the sign of both the lags and $\Delta\nu_{0}$, \citet{Eijnden2016} proposed a differential precession model. In their model, they assume that the QPO is caused by Lense-Thirring (LT) precession of the inner flow \citep{Ingram2009}; the inner flow is inhomogeneous, but consists of two separate halves which precess at different frequencies, with the inner half always precessing faster and hence producing a higher QPO frequency, as expected from LT precession. When the QPO frequency is above 2 Hz, the inner half spectrum is harder than the outer half, thus a QPO frequency that increases with energy is expected; when the QPO frequency is below 2 Hz, the spectral shape of the two halves switches so that the inner half has a softer spectrum than the outer half, thus the QPO frequency decreases with energy. Since the outer half will lag behind the inner half on average, soft lags will be observed when the inner half spectrum is harder, and vice versa. The main obstacle to this scenario is that no clear observational evidence exists for the required spectral change. Several possible scenarios to try and explain this contradiction have been discussed by \citet{Eijnden2016}.

Alternatively, the hard and soft QPO lags could originate from two completely different physical mechanisms. In GX 339$-$4, \citet{Zhang2017} found that the evolution of the lags of the type-C QPO with QPO frequency is similar to that of the reflected flux: both the QPO lags and the reflected flux increase first with QPO frequency, and then turn to flat when the QPO frequency is  $\sim1.7$ Hz. \citet{Zhang2017} found clear reflection features in the lag-energy spectra when the QPO frequency is above $\sim1.7$ Hz, suggesting that the reflection component contributes significantly to these lags. The absence of such features when the QPO freqeuncy is $\sim1.7$ Hz provides evidence that in GX 339$-$4, at QPO frequencies below $\sim1.7$ Hz, the QPO lags may be dominated by another mechanism. In GX 339$-$4, a dip-like feature at the QPO fundamental is only observed \citep{Zhang2017} in the lag-frequency spectra when the reflected flux is high at QPO frequencies above $\sim1.7$ Hz, suggesting that this feature may be related to the reflection process. \citet{Kotov2001} also found that the reflection component might have a suppressing effect on the observed lags that cause such features. Similar to GX 339$-$4, here we find a dip-like feature in the lag-frequency spectrum at the frequency of the QPO fundamental when the QPO frequency is above 2 Hz in GRS 1915+105. However, when the QPO frequency is below 2 Hz, no such dip-like feature is found; instead, in some cases, when the QPO frequency is below 0.5 Hz, a peak-like feature is observed in the lag-frequency spectrum at the frequency of the QPO fundamental. If the dip-like feature is indeed due to reflection, we would expect that the QPO lags below and above 2 Hz might originate from two different mechanisms. 

In Fig. \ref{fig:phase_rms}, we showed that the relation between the QPO fractional rms and the QPO average lags can be well fitted with a broken line: as the QPO average lags go from negative to positive, the QPO fractional rms first increases, reaching its maximum at zero lag, and then decreases. A possible explanation is that when the average phase lag between two broad bands at the QPO frequency is zero, the full-band light curve at that frequency would be the superposition of the two broad-band light curves in phase, yielding a large QPO rms. On the contrary, if the average phase lag is non-zero, the full-band light curve at the QPO frequency would be the superposition of the two broad-band light curves at different phases, yielding a smaller QPO rms. The fractional rms amplitude of a signal that is the superposition of two sinusoidal functions with the same amplitude $A$ shifted by a phase $\phi$ is ${\rm rms} \sim A \sqrt{{\rm cos}\phi+1}$. If $\phi = 0$, then ${\rm rms}(\phi = 0) \sim \sqrt{2}A$. Therefore, the rms will be suppressed by a factor $\sqrt{({\rm cos}\phi+1)/2}$ due to a non-zero lag between different energy bands. For lags of less than 0.4 rad as observed here, the effect on the rms is less than 3\% of the value of the QPO amplitude, whereas, we found that the changes are $\sim70\%$. This suggests that this mechanism is not responsible for the rms variations. 

\subsection{Sub/second-harmonic Lags}

We found that the phase-lag behaviour of the subharmonic of the type-C QPO in \target is quite similar to that of the QPO fundamental, with the subharmonic average lags decreasing with subharmonic frequency and change sign when the subharmonic frequency is around 1 Hz. The frequency at which the subharmonic average lags change sign is consistent with being half of the frequency at which the QPO fundamental average lags change sign. Previous work on other BHBs found both hard and soft subharmonic lags \citep[e.g.][]{Casella2005,Eijnden2017}. The subharmonic lags also show differences between high- and low- inclination sources, although the dependence upon the inclination is weaker than that of the QPO fundamental lags \citep{Eijnden2017}. The similar phase-lag behaviour between the fundamental and subharmonic of the type-C QPO suggests that these two QPO components may be produced by the same mechanism. 

On the contrary, the phase-lag behaviour of the second-harmonic of the type-C QPOs in \target is completely different than that of the QPO fundamental and subharmonic. The second harmonic of the QPO only shows hard lags that remain more or less constant as a function of the second-harmonic frequency, without changing sign at any particular frequency. This suggests that the second-harmonic lags may originate from a different mechanism (or region) than the lags of the QPO fundamental and subharmonic. Using frequency-resolved spectroscopy, \citet{Axelsson2016} found that the energy spectrum of the second harmonic of the type-C QPO in GX 339$-$4 is softer than the energy spectrum at the QPO fundamental, and the time-averaged spectrum. \citet{Axelsson2016} suggest that the Comptonization region that produces the QPO is inhomogeneous, whereas the second harmonic comes from the outer part of the flow, which would have a softer spectrum. An alternative interpretation was proposed by \citet{Eijnden2017}: the precessing inner flow irridating the disc on both the front and the back sides could produce the second harmonic. In this scenario, the lag is produced by the red- and blueshifting of the reflected harmonic.

\subsection{Slope of the Lag-energy Spectrum}

The relation between the QPO lags and photon energy for the type-C QPOs in \target can be well fitted by a log-linear function, as previously reported \citep{Reig2000,Qu2010,Pahari2013}. Such an energy dependence of the lags at the QPO frequency has also been found in other BHBs \citep{Wijnands1999,Zhang2017,Eijnden2017}. Here we found that the same energy dependence also holds for the subharmonic and second harmonic of the type-C QPOs in GRS 1915+105. In addition, we found that the slope of the lag-energy spectrum at the QPO frequency versus QPO frequency follows the same trend as the QPO average lags versus QPO frequency; this is also the case for the subharmonic and second harmonic of the QPO. In other words, the change of the average phase lags reflect a change of the lag-energy spectrum. 

For the fundamental and subharmonic of the type-C QPO in GRS 1915+105, as their average phase lags decrease and change sign from positive to negative, the slope of their lag-energy spectrum also decreases and changes sign at the same frequency. This provides evidence that the lag-energy spectrum simply pivots around an energy. Although the second harmonic shows a very different phase-lag behaviour, the average phase lags and the lag-energy spectrum slope of the second harmonic of the QPO follow the same trend. This indicates that the lag-energy spectrum drives the average lags of the QPO, and that this is the case for the QPO fundamental and both its subharmonic and second harmonic.

\section*{Acknowledgements}

The authors thank the anonymous referee for the useful comments that improved the paper. L.Z. acknowledges support from the Royal Society Newton Funds. D.A. acknowledges support from the Royal Society. L.Z., J.L.Q., L.C., Q.C.B, Y.H., X.M. and L.T. thank supports from the National Program on Key Research and Development Project (Grant No. 2016YFA0400803), and the National Natural Science Foundation of China (Grant No. 11673023, U1838201, U1838111, U1838115, 11733009 and U1838108). T.M.B. acknowledges financial contribution from the agreement ASI-INAF n. 2017-14-H.0.
This research has made use of data and/or software provided by the High Energy Astrophysics Science Archive Research Center (HEASARC), which is a service of the Astrophysics Science Division at NASA/GSFC and the High Energy Astrophysics Division of the Smithsonian Astrophysical Observatory.





\bsp	
\label{lastpage}
\end{document}